\documentclass[11pt]{article}
\pdfoutput=1

\usepackage{aas_macros}

\usepackage{cite}
\usepackage{booktabs}
\usepackage{amsmath,amssymb,amsbsy,amstext, amsthm, simplewick}
\usepackage{hyperref}
\usepackage{graphicx}
\usepackage{amsfonts}
\usepackage{amssymb}
\usepackage[small]{caption}
\usepackage{upgreek}
\usepackage[svgnames,dvipsnames,x11names,table]{xcolor}
\usepackage{multirow} 
\usepackage{geometry}
\usepackage[hang,flushmargin]{footmisc}
\usepackage{bm}
\usepackage{braket}
\usepackage{subcaption}
\usepackage{simpler-wick}
\usepackage{mathtools}

\usepackage{colortbl}
\makeatletter
\newlength{\apb@width}
\newcommand{\autoparbox}[2][c]{\settowidth{\apb@width}{#2}\parbox[#1]{\apb@width}{#2}}

\makeatother

\definecolor{lightgray}{gray}{0.9}

\usepackage[framemethod=default]{mdframed}
\newmdenv[skipabove=7pt,
skipbelow=7pt,
rightline=false,
leftline=false,
topline=false,
bottomline=false,
backgroundcolor=gray!10,
linecolor=gray,
innerleftmargin=5pt,
innerrightmargin=5pt,
innertopmargin=5pt,
innerbottommargin=5pt,
leftmargin=0cm,
rightmargin=0cm,
linewidth=4pt]{eBox}

\numberwithin{equation}{section}

\def\beq{\begin{equation}}
\def\eeq{\end{equation}}

\def\bea{\begin{eqnarray}}
\def\eea{\end{eqnarray}}

\def\Neff{N_{\rm eff}}
\def\beq{\begin{equation}}
\def\eeq{\end{equation}}
\def\bea{\begin{eqnarray}}
\def\eea{\end{eqnarray}}

\def\Mpl{M_{\rm pl}}

\def\fnlloc{f^{\rm loc}_{\rm NL}}

\def\k{{\vec k}}

\def\x{{\vec x}}

\DeclareRobustCommand{\SkipTocEntry}[4]{}

\def\@fnsymbol#1{\ensuremath{\ifcase#1\or $^\clubsuit$\or \dagger\or \ddagger\or
   \mathsection\or \mathparagraph\or \|\or **\or \dagger\dagger
   \or \ddagger\ddagger \else\@ctrerr\fi}}

\definecolor{blue3}{RGB}{31, 119, 180}
\definecolor{red3}{RGB}{	214, 39, 40}
\definecolor{orange3}{RGB}{255, 127, 14}
\definecolor{green3}{RGB}{44, 160, 44}

\usepackage{lineno}
\begin{document}

%\begin{titlepage}
\setcounter{page}{1} \baselineskip=15.5pt 
\thispagestyle{empty}

\begin{center}
{\fontsize{18}{18} \bf Snowmass Theory Frontier: \\[5pt]  Astrophysics and Cosmology }
\end{center}

\vskip 20pt
\begin{center}
\noindent
{\fontsize{12}{18}\selectfont  Daniel Green$^\clubsuit$\footnote{ \href{mailto:drgreen@physics.ucsd.edu}{drgreen@physics.ucsd.edu}  }, Joshua T. Ruderman$^\diamondsuit$\footnote{ \href{mailto:ruderman@nyu.edu}{ruderman@nyu.edu}  }, Benjamin R. Safdi$^{\spadesuit}$\footnote{ \href{mailto:brsafdi@berkeley.edu}{brsafdi@berkeley.edu}  }, Jessie Shelton$^\heartsuit$\footnote{ \href{mailto:sheltonj@illinois.edu }{sheltonj@illinois.edu }  }}
\end{center}

\begin{center}
\textit{ $^\clubsuit$ Department of Physics, University of California San Diego, La Jolla, CA 92093, USA}\\
\textit{$^\diamondsuit$ Center for Cosmology and Particle Physics, Department of Physics, New York University, New York, NY 10003, USA}\\
\textit{$^\spadesuit$ Department of Physics,  University  of  California,  Berkeley,  CA  94720,  USA}\\
\textit{Physics Division,  Lawrence  Berkeley  National  Laboratory,  Berkeley,  CA  94720,  USA} \\
\textit{$^\heartsuit$ Illinois Center for Advanced Studies of the Universe, Department of Physics, University of Illinois Urbana-Champaign, Urbana, IL 61801, USA }
\end{center}

\vskip 20pt
\begin{center}
\noindent
%{\fontsize{12}{18}\selectfont 
{Ana Ach\'ucarro$^{1}$, Peter Adshead$^2$, Yashar Akrami$^{3,4}$, Masha Baryakhtar$^{5}$, Daniel Baumann$^{6,7}$, Asher Berlin$^{8}$, Nikita Blinov$^{9}$, Kimberly K. Boddy$^{10}$, Malte Buschmann$^{11}$, Giovanni Cabass$^{12}$,  Robert Caldwell$^{13}$, Emanuele Castorina$^{14}$, Thomas Y. Chen$^{15}$, Xingang Chen$^{16}$,  William Coulton$^{17}$, Djuna Croon$^{18}$, Yanou Cui$^{19}$, David Curtin$^{20}$, Francis-Yan Cyr-Racine$^{21}$, Christopher Dessert$^{22}$, Keith R. Dienes$^{23,24}$, Patrick Draper$^{2}$, Peizhi Du$^{25}$, Sebastian A. R. Ellis$^{26}$, Rouven Essig$^{27}$, Raphael Flauger$^{28}$, Chee Sheng Fong$^{29}$, Joshua W. Foster$^{30}$,  Jacopo Fumagalli$^{31, 32}$, Keisuke Harigaya$^{33,34,35}$, Shunsaku Horiuchi$^{36}$, Mikhail M. Ivanov$^{12}$, Yonatan Kahn$^{2}$, Simon Knapen$^{37}$,  Rebecca K. Leane$^{38}$, Hayden Lee$^{35}$, Erik W. Lentz$^{39}$, Matthew Lewandowski$^{40}$, Mariangela Lisanti$^{11}$, Andrew J. Long$^{41}$, Marilena Loverde$^{5}$, Azadeh Maleknejad$^{42}$, Liam McAllister$^{43}$,  Samuel D. McDermott$^{8}$, Robert McGehee$^{44}$, P. Daniel Meerburg$^{45, 46}$, Joel Meyers$^{47}$, Azadeh Moradinezhad Dizgah$^{26}$, Moritz M\"unchmeyer$^{48}$, Nadav Joseph Outmezguine$^{37,49}$, Enrico Pajer$^{50}$, Gonzalo A. Palma$^{51}$, Aditya Parikh$^{52}$, Jong-Chul Park$^{53}$, Annika H. G. Peter$^{54}$, Guilherme L. Pimentel$^{55}$, S\'ebastien Renaux-Petel$^{56}$, Nicholas L. Rodd$^{42}$, Bibhushan Shakya$^{57}$, Gary Shiu$^{48}$, Eva Silverstein$^{58}$, Marko Simonovic$^{42}$, Rajeev Singh$^{59}$, Charlotte Sleight$^{60}$, Volodymyr Takhistov$^{61}$,  Philip Tanedo$^{19}$, Massimo Taronna$^{62,63}$,  Brooks Thomas$^{64}$, Natalia Toro$^{38}$, Yu-Dai Tsai$^{65}$, Edoardo Vitagliano$^{66}$, Mark Vogelsberger$^{67}$, Benjamin Wallisch$^{28,12}$, Benjamin D. Wandelt$^{56,17}$, Risa H. Wechsler$^{68,58,38}$, Christoph Weniger$^{6}$, W. L. Kimmy Wu$^{37}$, Weishuang Linda Xu$^{37,49}$, Masaki Yamada$^{69}$, Hai-Bo Yu$^{19}$, Zhengkang Zhang$^{70}$, Yi-Ming Zhong$^{35}$, Kathryn Zurek$^{71}$ }
\end{center}

\begin{center}
\textit{$^{1}$ Lorentz Institute for Theoretical Physics, Leiden University, 2333 CA Leiden, The Netherlands} \\
\textit{$^{2}$ Illinois Center for Advanced Studies of the Universe, Department of Physics, University of Illinois Urbana-Champaign, Urbana, IL 61801, USA} \\    
\textit{$^{3}$ CERCA/ISO, Department of Physics, Case Western Reserve University, 10900 Euclid Avenue, Cleveland, OH 44106, USA}\\
\textit{$^{4}$ Department of Physics, Imperial College London, Blackett Laboratory, Prince Consort Road, London SW7 2AZ, United Kingdom} \\    
\textit{$^{5}$ Department of Physics, University of Washington, Seattle, WA 98195, USA} \\ 
\textit{$^{6}$ Center for Theoretical Physics, National Taiwan University, Taipei 10617, Taiwan} \\
\textit{$^7$ Institute of Physics, University of Amsterdam, Amsterdam, 1098 XH, The Netherlands} 
\\    
\textit{$^{8}$ Fermi National Accelerator Laboratory, Theory Division, Batavia, Illinois, 60510, USA} \\    
\textit{$^{9}$ Department of Physics and Astronomy, University of Victoria, Victoria, BC, Canada} \\    
\textit{$^{10}$ Department of Physics, The University of Texas at Austin, Austin, TX 78712, USA} \\ 
\textit{$^{11}$ Department of Physics, Princeton University, Princeton, NJ 08544, USA} \\  
\textit{$^{12}$ School of Natural Sciences, Institute for Advanced Study, Princeton, NJ 08540, USA} \\
\textit{$^{13}$ Department of Physics and Astronomy, Dartmouth College, 6127 Wilder Laboratory, Hanover, NH 03755 USA} \\ 
\textit{$^{14}$ Department of Physics ``Aldo Pontremoli'', University of Milan, Milan, Italy} \\
\textit{$^{15}$ Fu Foundation School of Engineering and Applied Science, Columbia University, New York, NY 10027, USA} \\
\textit{$^{16}$ Institute for Theory and Computation, Harvard-Smithsonian Center for Astrophysics, 60 Garden Street, Cambridge, MA 02138, USA} \\ 
\textit{$^{17}$ Center for Computational Astrophysics, Flatiron Institute, 162 5th Avenue, New York, NY 10010, USA} \\
\textit{$^{18}$ Institute for Particle Physics Phenomenology, Durham University, Durham DH1 3LE, UK} \\ 
\textit{$^{19}$ Department of Physics and Astronomy, University of California, Riverside, CA 92521, USA} \\
\textit{$^{20}$ University of Toronto, Toronto, Canada} \\
\textit{$^{21}$ Department of Physics and Astronomy, University of New Mexico, Albuquerque, NM 87106, USA}\\
\textit{$^{22}$ Center for Cosmology and Particle Physics, Department of Physics, New York University, New York, NY 10003, USA} \\    
\textit{$^{23}$ Department of Physics, University of Arizona, Tucson, AZ  85721  USA} \\
\textit{$^{24}$ Department of Physics, University of Maryland, College Park, MD 20742, USA} \\
\textit{$^{25}$ C.N.~Yang Institute for Theoretical Physics, Stony Brook University, NY 11794, USA} \\
\textit{$^{26}$ D\'epartement de Physique Th\'eorique, Universit\'e de Gen\`eve, 24 quai Ernest Ansermet, 1211 Geneva 4, Switzerland}  \\       
\textit{$^{27}$ C.N.~Yang Institute for Theoretical Physics, Stony Brook University, NY 11794, USA} \\   
\textit{$^{28}$ Department of Physics, University of California San Diego, La Jolla, CA 92093, USA} \\
\textit{$^{29}$ Centro de Ci\^encias Naturais e Humanas, Universidade Federal do ABC, 09.210-170, Santo Andr\'e, SP, Brazil} \\
\textit{$^{30}$ Center for Theoretical Physics, Massachusetts Institute of Technology, Cambridge, MA 02139, USA} \\  
\textit{$^{31}$ Instituto de F\'{ı}sica Te\'{o}rica UAM/CSIC, Calle Nicol\'as Cabrera 13-15, Cantoblanco E-28049 Madrid, Spain}\\
\textit{$^{32}$ Departamento de F\'{ı}sica Te\'{o}rica, Universidad Aut\'{o}noma de Madrid (UAM), Campus de Cantoblanco, E-28049 Madrid, Spain}\\
\textit{$^{33}$ Department of Physics, University of Chicago, Chicago, IL 60637, USA
} \\
\textit{$^{34}$ Enrico Fermi Institute, University of Chicago, Chicago, IL 60637, USA} \\
\textit{$^{35}$ Kavli Institute for Cosmological Physics, University of Chicago, Chicago, IL 60637, USA} 
\\    
\textit{$^{36}$ Center for Neutrino Physics, Department of Physics, Virginia Tech, Blacksburg, Virginia 24061, USA} 
\\    
\textit{$^{37}$ Physics Division,  Lawrence  Berkeley  National  Laboratory,  Berkeley,  CA  94720,  USA} \\
\textit{$^{38}$ SLAC National Accelerator Laboratory, Stanford University, Stanford, CA 94039, USA} \\
\textit{$^{39}$ Pacific Northwest National Laboratory, Richland, WA 99354, USA} \\     
\textit{$^{40}$Department of Physics and Astronomy, Northwestern University, Evanston, IL 60208 USA} \\
\textit{$^{41}$ Department of Physics and Astronomy, Rice University, Houston, TX 77005, USA} \\
\textit{$^{42}$ Theoretical Physics Department, CERN, 1 Esplanade des Particules, CH-1211 Geneva 23, Switzerland}  \\  
\textit{$^{43}$ Department of Physics, Cornell University, Ithaca, NY 14853, USA}  \\  
\textit{$^{44}$ Leinweber Center for Theoretical Physics, Department of Physics, University of Michigan, Ann Arbor, MI 48109, USA}  \\
\textit{$^{45}$ Van Swinderen Institute for Particle Physics and Gravity, University of Groningen, Nijenborgh 4, 9747 AG Groningen, The Netherlands}  \\
\textit{$^{46}$ 
Kapteyn Astronomical Institute, University of Groningen, 9700 AV Groningen, The Netherlands}  \\   
\textit{$^{47}$ Department of Physics, Southern Methodist University, Dallas, TX 75275, USA} \\    
\textit{$^{48}$ Department of Physics, University of Wisconsin-Madison, Madison, WI 53706, USA} \\   
\textit{$^{49}$ Department of Physics,  University  of  California,  Berkeley,  CA  94720,  USA} \\  
\textit{$^{50}$ Department of Applied Mathematics and Theoretical Physics, University of Cambridge, Wilberforce Road, Cambridge, CB3 0WA, UK} \\  
\textit{$^{51}$ Departamento de F\'isica, Facultad de Ciencias F\'isicas y Matem\'aticas, Universidad de Chile, Santiago, Chile } \\
\textit{$^{52}$ Department of Physics, Harvard University, Cambridge, MA 02138, USA}\\
\textit{$^{53}$ Department of Physics, Chungnam National University, Daejeon 34134, Republic of Korea}\\
\textit{$^{54}$   CCAPP, Department of Physics, and Department of Astronomy, The Ohio State University, Columbus, OH 43210 USA } \\
\textit{$^{55}$ Scuola Normale Superiore and INFN, Piazza dei Cavalieri 7, 56126 Pisa, Italy} \\  
\textit{$^{56}$ Institut d'Astrophysique de Paris, CNRS et Sorbonne Universit\'e, 98 bis bvd Arago 75014 Paris, France} \\    
\textit{$^{57}$ Deutsches Elektronen-Synchrotron DESY, Notkestr. 85, 22607 Hamburg, Germany} \\ 
\textit{$^{58}$ Department of Physics, Stanford University, Stanford, CA 94305 USA} \\
\textit{$^{59}$ Institute of Nuclear Physics Polish Academy of Sciences, PL-31-342, Krak\'{o}w, Poland} \\  
\textit{$^{60}$ Centre for Particle Theory and Department of Mathematical Sciences, Durham University, Durham, DH1 3LE, UK} \\ 
\textit{$^{61}$ International Center for Quantum-field Measurement Systems for Studies of the Universe and Particles (QUP), High Energy Accelerator Research Organization (KEK), Tsukuba, Ibaraki 305-0801, Japan} \\  
\textit{$^{62}$ Dipartimento di Fisica ``Ettore Pancini", Universit degli Studi di Napoli Federico II, Monte S. Angelo, Via Cintia, 80126 Napoli, Italy} 
\\  
\textit{$^{63}$ Scuola Superiore Meridionale, Largo San Marcellino 10, 80138 Napoli, Italy}  \\  
\textit{$^{64}$ Department of Physics, Lafayette College, Easton, PA 18042 USA} 
\\   
\textit{$^{65}$ Department of Physics \& Astronomy, University of California, Irvine,
Irvine, CA 92697, USA} \\
\textit{$^{66}$ Department of Physics and Astronomy, University of California, Los Angeles, 475 Portola Plaza, Los Angeles, CA 90095-1547, USA} \\ 
\textit{$^{67}$ Department of Physics and Kavli Institute for Astrophysics and Space Research, Massachusetts Institute of Technology, Cambridge, MA 02139, USA}   \\
\textit{$^{68}$ Kavli Institute for Particle Astrophysics and Cosmology, Stanford University, Stanford, CA 94305, USA}   \\
\textit{$^{69}$ Department of Physics, Tohoku University, Sendai, Miyagi 980-8578, Japan} \\ 
\textit{$^{70}$ Department of Physics, University of California, Santa Barbara, CA 93106, USA} \\ 
\textit{$^{71}$ Walter Burke Institute for Theoretical Physics, California Institute of Technology, Pasadena, CA, USA}\\
\end{center}

%=========================================
\vspace{0.4cm}
 \begin{center}{\bf Abstract}
  \end{center}
   We summarize progress made in theoretical astrophysics and cosmology over the past decade and areas of interest for the coming decade.  This Report is prepared as the TF09 ``Astrophysics and Cosmology" topical group summary for the Theory Frontier as part of the Snowmass 2021 process.

%\noindent

%\end{titlepage}

%\restoregeometry

%\newpage

\newpage

\tableofcontents

\newpage

\section{Executive Summary}

Many of the most fundamental questions in physics are tied to our understanding of the universe on astrophysical and cosmological scales.
In fact, apart from neutrino mass, which itself was first discovered through astrophysical neutrinos, all unambiguous experimental evidence for physics beyond the known laws and particle content of nature comes from astrophysical and cosmological observations. 
Cosmology and astrophysics provide a wide range of opportunities to expand our knowledge of the fundamental laws of nature, both through direct searches for beyond-the-Standard-Model (BSM) physics and through tests of the Standard Model (SM) in extreme conditions that are impossible to recreate in the laboratory.  The data acquired to date have had profound ramifications across virtually all domains of theoretical particle physics, including string theory and quantum gravity, particle phenomenology and model-building, amplitudes, effective field theory (EFT), lattice quantum chromodynamics (QCD), neutrino physics, and beyond. These theoretical developments have, in turn, become the driving factor governing the next generation of experiments searching for new physics in astrophysical and cosmological contexts.  

Broadly, the impact of theoretical effort in cosmology and astrophysics over the past decade can be viewed through the lenses of (i) advancing our understanding of fundamental physics by forcing us to ponder extreme scenarios where e.g.~quantum effects and gravity must be considered simultaneously, (ii) developing new microscopic models that can potentially explain the outstanding problems facing our understanding of nature, and (iii) inventing new approaches  
to test our best-motivated theories, in addition to developing the theoretical tools needed to properly interpret the resulting data.    
A decade ago dark matter (DM) was being searched for in limited parts of parameter space; today, because of theoretical work that expanded the range of viable DM models and proposed detection concepts for leading candidates, a large ecosystem of experimental approaches to DM may now put us on the edge of discovery.

All of the leading DM models today, in addition to almost all of the major DM experimental projects, were proposed in theoretical works.  
Novel analyses of astrophysical data in searches for DM have unveiled anomalies  
and helped resolve outstanding observations of the cosmos. At the same time, a whole host of new astrophysical and cosmological signatures of new physics in the early universe have emerged with the expanded development of theories of non-minimal dark particle physics.  
Observational tests of the inflationary epoch have been driven by deepening our understanding of the range of possible signals and astrophysical foregrounds.  These theoretical developments will be critical for achieving the observational goals in the coming generation of cosmic and astrophysical surveys. In addition, progress in our understanding of cosmology (including e.g.~inflation, dark energy, and large scale structure) has been spurred by progress in more formal theoretical techniques (including e.g.~quantum gravity, the conformal bootstrap, amplitudes and EFT) and inspires new directions of theoretical investigation across the frontier.

In summary, theoretical work in astrophysics and cosmology over the past decade has played a crucial role in informing our understanding of theoretical physics, developing current and future projects, and innovating the tools needed to make sure these experiments are able to successfully extract meaningful and transformative science results.

This report summarizes the work done in the topical group TF09 and draws on the Snowmass2021 white papers~\cite{Adams:2022pbo,Blinov:2022tfy,Baryakhtar:2022obg,Achucarro:2022qrl,Amin:2022soj,Cabass:2022avo,Chang:2022tzj,CMB-HD:2022bsz,Abazajian:2022nyh,Dvorkin:2022jyg,
Dienes:2022zbh,Asadi:2022njl,Antypas:2022asj,Alvarez:2022kut,Ghosh:2022zef,Ackermann:2022rqc,Mitridate:2022tnv,Berti:2022wzk,
Dawson:2022oig,Ferraro:2022cmj,Karkare:2022bai,Liu:2022iyy,Bechtol:2022koa,Alvarez:2022rbk,Flauger:2022hie,Baumann:2022jpr,deRham:2022hpx,Draper:2022pvk,Hartman:2022zik,Kruczenski:2022lot,Agrawal:2022rqd,Bousso:2022ntt,Leane:2022bfm,Carney:2022gse,Faulkner:2022mlp,Brito:2022lmd,Bird:2022wvk,Elor:2022hpa,Banerjee:2022qcb,Essig:2022yzw,Boddy:2022knd,Berlin:2022hfx,Craig:2022uua,Caldwell:2022qsj,Aramaki:2022zpw}.

\section{Topical Introduction}

\subsection{Dark Matter}

Interpretations of cosmic microwave background (CMB) data point to $\sim$27\% of the energy density in our universe today being in the form of DM, with $\sim$5\% in baryons and the rest in dark energy~\cite{Planck:2018vyg}.  The evidence for DM in the CMB data is overwhelming.
Density perturbations generated in the matter density from quantum fluctuations during inflation are now understood to have collapsed to form DM halos, which seeded galaxy clusters and galaxies like our own Milky Way.  Thanks in large part to recent theoretical work on $N$-body simulations and analytic methods, the comparison of the predicted DM structure in the standard cold DM plus cosmological constant paradigm ($\Lambda$CDM) to measured large scale structure (LSS) and halo data is now a precision science~\cite{Banerjee:2022qcb}.  Equally compelling data for the existence of cold and at most weakly interacting DM across a variety of astrophysical scales arises from galaxy rotation curves~\cite{Salucci:2018hqu,1978ApJ...225L.107R,1980ApJ...238..471R}, stellar kinematics~\cite{Bovy:2012tw}, and weak and strong gravitational lensing~\cite{Mandelbaum:2005nx,2012ApJ...757...82B,Natarajan:2017sbo}.  It is not possible to explain the observed DM within the SM of particle physics; DM requires the existence of at least one new fundamental particle of nature, though DM itself may be composite.  The singular possible exception to this is if DM is made of primordial black holes, but (i) these have become increasingly constrained as DM over the past decade, and (ii) even if DM were primordial black holes, new fundamental dynamics would be needed to source the large density perturbations in the early universe needed for their creation (see~\cite{Bird:2022wvk} for a Snowmass2021 discussion of primordial black hole DM).

The DM mass is constrained on the lower end to be larger than $\sim$$10^{-20}$ eV and on the upper end to be smaller than $\sim$$10^4$ $M_\odot$.  The lower bound comes from the De Broglie wavelength ($\lambda_{\rm DB} \sim {\rm eV} / m_{\rm DM}$ mm) of the DM being galactic scale for DM masses $m_{\rm DM} \sim 10^{-22}$ eV and typical galactic velocities.  Such ultralight DM candidates are referred to as fuzzy DM, and this paradigm has received a significant amount of attention over the past decade because of ``top-down" motivations in the context of string theory, data-driven observations that fuzzy DM could alleviate tensions in some structure formation anomalies, and the plethora of relevant new astrophysical data and theoretical modeling and simulation techniques that have been gathered and developed in recent years.  We review fuzzy DM further in Sec.~\ref{sec:axions}.  On the upper end of the mass spectrum, sufficiently compact DM with masses above $\sim$$10^3$-$10^4$ $M_\odot$ would be too granular on astrophysical scales; for example, DM masses above $\sim$$10^3$ $M_\odot$ have been suggested to lead to the disruption of globular clusters~\cite{1985ApJ...299..633L,1993ApJ...413L..93M,1993ApJ...417..220G}, while DM masses above $\sim$$10^4$ $M_\odot$ would likely inject too much Poisson noise in the Ly$\alpha$ forest~\cite{Afshordi:2003zb}.

We find ourselves at a unique point in history, where the existence of DM on astrophysical and cosmological scales is known and well characterized, but the microscopic nature of the DM is almost completely unconstrained, up to rough constraints on the DM mass and interaction strengths with itself and with ordinary matter.  However, over the past decades theorists have developed particle-physics-based models to explain DM, and typically these models lead to faint but observable signatures.  Yet, the types of signatures vary drastically depending on the DM model at hand, motivating a world-wide scientific program searching for evidence of particle DM across laboratory experiments ranging from the Large Hadron Collider (LHC) to underground direct detection experiments to precision laboratory experiments.  On the other hand, in many DM models there are direct or indirect astrophysical signatures, such as modifications to DM halo structure itself ({i.e.}, direct modifications to the assumptions in $\Lambda$CDM), electromagnetic, cosmic ray, and neutrino signatures in the case of annihilating, decaying, and converting DM, and even gravitational wave signatures from DM associated with phase transitions in the early universe.  DM may also be associated with new light physics, such as light mediators or ultralight particles, that may have effects ranging from modifying stellar evolution to producing signatures in dedicated fixed target experiments to adding extra radiation contributing to the effective number of neutrino degrees of freedom $N_{\rm eff}$ measured in the early universe from the CMB.  The theoretical efforts to produce viable DM candidates have been the driving force behind numerous experimental and observational programs aimed at finding evidence for these signatures.  Beyond the construction of specific DM models, in the past decade theorists have played an essential role in proposing and helping implement specific experiments and astrophysical search strategies for covering some of the best motivated DM parameter space, pushing science closer to a transformative discovery.    

In addition to being one of the central driving forces for new experiments and projects across the cosmic, energy, and precision frontiers, theoretical efforts related to DM connect across the theory frontier.  From a bottom-up perspective many DM models address other outstanding problems in the SM\null. Weakly-interacting massive particle (WIMP) DM may arise in models that address the hierarchy problem related to the unnaturally light Higgs mass; the hierarchy problem has become even more striking over the past decade in light of the discovery of the Higgs boson and null results for searches for new physics at the LHC\null.  However, we argue in Sec.~\ref{sec:thermal} that WIMP DM, with mass near the electroweak scale, in addition to closely related variants that have been explored recently, remain promising DM candidates.  Theoretical work has been crucial in identifying promising thermal DM candidates, such as {e.g.} the supersymmetric higgsino, and calculating the signatures of these candidates in the laboratory and in astrophysics.  The QCD axion, which was originally introduced to solve the strong-CP problem related to the neutron electric dipole moment (EDM) but which may also explain the observed DM, is intimately connected to QCD, including lattice QCD, and the precision frontier, as we overview in Sec.~\ref{sec:axions}.  At the same time, axions are now known to emerge generically in the context of ``top-down" constructions such as those based on string theory.  Axions may be connected to deep concepts in quantum gravity, such as extra dimensions and the weak gravity conjecture, and these connections have been sharpened over the past decade. In particular, the QCD axion and axion-like particles, if they exist, may possibly be the only way of experimentally probing string theory compactifications, if string theory is indeed realized in our universe.  Axions may also be responsible for inflation and, furthermore, are now one of the leading motivations for searching for new-physics contributions to $N_{\rm eff}$ with next-generation CMB and LSS experiments and analyses.  

Sterile neutrino DM is naturally intertwined with the neutrino frontier, as $\sim$keV-scale sterile neutrinos would be connected to the active neutrinos through {e.g.}~the see-saw mechanism.  On the other hand sterile neutrinos are also a warm DM candidate, meaning that their finite temperature may wash out structure on small astrophysical scales.  Sterile neutrino DM, along with other DM candidates such as self-interacting DM and fuzzy DM, have been one of the driving forces for developing increasingly more sophisticated $N$-body simulations of structure formation in the early and late universe, since these models predict small-scale modifications relative to $\Lambda$CDM that may be detectable in galaxy and halo data.  Certain self-interacting DM candidates, such as glueball DM, are also motivated by string theory constructions that predict dark gauge groups.  More generally, the concept of ``dark sectors" has undergone a revolution over the past decade.  The idea of dark sectors is that DM may be part of a hidden sector, which has no constituents charged under the SM gauge groups but rather communicates to the SM through a number of possible ``portals."  Many explicit dark-sector constructions favor the keV-GeV DM mass range, which has led to a
burgeoning experimental program pushing towards direct detection experiments with lower thresholds and increasingly sensitive accelerator-based DM searches.
This program has been led by theorists and requires, in many cases, interdisciplinary work between {e.g.}~condensed matter theory and particle theory, in conjunction with cutting-edge experimental techniques such as quantum sensors.

\subsection{Cosmic History and Beyond}

The history of the universe spans many epochs.  Cosmological observations are sensitive to the period beginning with inflation and continuing through reheating, radiation-, matter-, and dark energy-domination.  Throughout this history, the energy and distance scales relevant to cosmic surveys span from the Planck scale to the size of the observable universe.  The study of our cosmic history unifies these disparate scales and brings together insights from every corner of theoretical physics~\cite{Flauger:2022hie}.  

At this moment, our understanding of the universe is challenged by both deep theoretical challenges and tensions in observations. Yet, the theoretical machinery we have constructed gives remarkable agreement with our most precise cosmic probes.  Perhaps the most basic and important question in cosmology (or even physics) today is why this machinery should work so well.  Dark energy dominates the universe today, and yet we have failed to understand its small size and whether an explanation in terms of a cosmological constant is consistent with theories of quantum gravity~\cite{Draper:2022pvk}.  
Yet, even given the unexplained contents of the universe, like baryon number, DM, and dark energy, we also know there is a fundamental limit to our statistical predictions starting from inflation.  We observe a quantum mechanical universe from within, unable to separate the dynamics of spacetime from the measurements of an observer.  The formal manifestation of this problem is that we lack a rigorous definition of cosmological observables.  More conceptually, cosmology forces us to confront our understanding of quantum mechanics when there is no external observer. This basic problem underlies the many challenges of understanding cosmological spacetimes within theories of quantum gravity or using holography.

Fortunately, the answers to some of these questions are encoded in maps of the universe.  The program of measuring the distribution of matter in our observable volume continues, with the promise of pushing our understanding well beyond what we have learned from the CMB~\cite{Amin:2022soj}.  Rapidly improving maps of the polarized CMB will deepen our understanding of the recombination era and could reveal primordial gravitational waves~\cite{Achucarro:2022qrl} and/or new particles and forces~\cite{Dvorkin:2022jyg}.  In addition, the CMB is an increasingly powerful probe of the late universe, as a backlight for gravitational lensing and spectral distortions by the intervening matter.  Galaxy surveys and intensity mapping also directly measure the matter at low redshifts with statistical power that is already rivaling the CMB~\cite{Ferraro:2022cmj}.

As the frontier of cosmic surveys pushes into these new domains, our ability to answer our deepest questions about the universe is increasingly limited by our theoretical tools to understand the data itself~\cite{Amin:2022soj}.  Our measurement of primordial gravitational waves from the CMB will be limited by our ability to remove the effects of dust and gravitional lensing from the maps of the B-mode polarization.  Maps of the low redshift distribution of matter are limited by our understanding of nonlinear evolution on short distances; vastly more information is available in the maps of the distribution of galaxies than we are able to use for inferring cosmological parameters.  Yet, these problems are not beyond the power of ingenuity of the theoretical physics community.  In the past decade, the tools needed to surpass these limitations have been demonstrated on real data.  In addition, theoretical insights are opening the door to entirely new methods for exploring our universe.  In the next decade, the continued investment in these efforts is expected to yield dramatic returns in the resulting scientific insights generated by cosmic surveys.

\subsection{Baryogenesis}

As is readily apparent, the universe is mostly matter and not anti-matter.  The present-day primordial baryon asymmetry is quantified precisely through big-bang nucleosynthesis (BBN) and CMB analyses to be $\eta_B = (n_b - n_{\bar b}) / n_\gamma = n_b / n_\gamma = (6.10 \pm 0.4)\times 10^{-10}$~\cite{Planck:2018vyg}.  The observed baryon asymmetry cannot be explained within the SM in conjunction with the standard $\Lambda$CDM cosmology; new physics is required, and the BSM production of the primordial baryon asymmetry is referred to as baryogenesis.   (See~\cite{Elor:2022hpa} and section 3 of~\cite{Asadi:2022njl} for discussions of recent ideas in baryogenesis.)  The development of new models for baryogenesis, and the exploration of their potential experimental tests, constitutes a vibrant area within the theory frontier.   Baryogenesis could have occurred at a wide variety of different energy scales, and therefore experimental signals of baryogenesis are model-dependent.  Possible signals include new sources of CP violation and cosmological imprints of non-equilibrium physics, such as the production of gravitational waves.

Classic scenarios for baryogenesis include leptogenesis~\cite{Fukugita:1986hr}, where a lepton asymmetry  is generated by right-handed neutrino decays and transferred to a baryon symmetry though electroweak sphalerons, and electroweak baryogenesis, where the baryon asymmetry is generated during the electroweak phase transition (for reviews see~\cite{Cline:2006ts, Morrissey:2012db, Konstandin:2013caa, Garbrecht:2018mrp, Bodeker:2020ghk}).  Recent work on leptogenesis and electroweak baryogenesis have explored experimental signals of these scenarios, including gravitational waves (see for example~\cite{Buchmuller:2013lra} and~\cite{Caprini:2019egz}).

Several more recent proposals for baryogenesis have emphasized possible connections between the baryon asymmetry and DM\null.  These include asymmetric DM (for reviews see~\cite{Davoudiasl:2012uw, Petraki:2013wwa, Zurek:2013wia}), WIMP baryogenesis~\cite{Cui:2011ab}, and axion baryogenesis~\cite{Co:2019wyp}.  Other recent proposals produce the baryon asymmetry from particle-anti-particle oscillations~\cite{Ipek:2016bpf} or from the interactions of SM mesons~\cite{Elor:2018twp}.

\section{Particle Dark Matter and Indirect Detection}
\label{sec:thermal}

\subsection{WIMP Dark Matter}

Weakly interacting massive particles (WIMPs) are DM candidates with electroweak (EW) scale masses and couplings that acquire their relic abundance through thermal freeze-out in the early universe (see~\cite{Bertone:2016nfn,Roszkowski:2017nbc} for modern reviews).  The excitement for laboratory-based WIMP searches was amplified by the fact that solutions to the hierarchy problem related to the unnaturally-low Higgs mass, such as electroweak-scale supersymmetry (SUSY), naturally produce WIMP-like DM candidates.  However, at present the search for WIMP DM is at a turning point.  Natural solutions to the hierarchy problem have been searched for directly at collider experiments, most recently with the LHC, and indirectly through a host of precision measurements.  No evidence for new weak-scale physics has been discovered to date, likely implying some degree of fine tuning in the electroweak sector (see~\cite{Craig:2022uua} for discussion). Furthermore,  searches for WIMP DM through missing energy at the LHC~\cite{Kahlhoefer:2017dnp}, scattering in underground direct detection experiments~\cite{Schumann:2019eaa}, and indirectly through astrophysical gamma-ray~\cite{Fermi-LAT:2016uux} and cosmic-ray~\cite{AMS:2019rhg} signatures, in addition to energy-injection in the early universe~\cite{Slatyer:2015jla}, tightly constrain the properties of a putative WIMP DM candidate (but see~\cite{Leane:2022bfm} for a discussion of possible astrophysical anomalies hinting towards WIMP DM).
Theorists have played a crucial role in interpreting the null observations from the ensemble of searches described above to tightly constrain 
natural solutions to the hierarchy problem and WIMP DM in its simplest forms~\cite{Roszkowski:2017nbc}. On the other hand, theorists have established that WIMP DM is very much not dead~\cite{Leane:2018kjk}.
Nearly-pure higgsino DM serves as an illustrative example of a surviving WIMP DM candidate that is as an exceptionally well-motivated target for upcoming experiments, as we now discuss.

Nearly-pure higgsino DM arises naturally in the context of top-down models such as 
Split SUSY~\cite{Arkani-Hamed:2004ymt}, where only the gauginos and higgsinos have masses near the TeV scale while the sfermions and gravitino have masses of order a large SUSY-breaking scale $\gtrsim$100 TeV\null. 
These models preserve gauge unification and may still accommodate neutralino DM, at the cost of introducing a sizable tuning between the EW scale and the SUSY-breaking scale.     
The higgsino may naturally obtain the correct relic abundance if it has a mass $1.0 \pm 0.1$ TeV~\cite{Cirelli:2005uq}. 
The two neutral higgsino states can naturally obtain a small mass splitting from mixing with other neutralino states.  A mass splitting $\delta m_N \approx 200$ keV is needed in order to prohibit $Z$-exchange for direct detection~\cite{Tucker-Smith:2001myb}; for splittings above this value the thermal higgsino direct detection cross-section is below the neutrino floor, meaning that future direct detection efforts for higgsino DM will be exceedingly difficult.  On the other hand, as has been shown in a series of theoretical works in recent years~\cite{Rinchiuso:2020skh,Dessert:2022evk}, indirect searches for higgsino DM annihilation into gamma-ray signatures will be conclusively tested in coming years with next-generation telescopes such as the Cherenkov Telescope Array (CTA).   
For fully characterizing the signals of annihilating TeV higgsino DM, more theoretical work is needed, building off of the body of work mostly developed in the last decade~\cite{Boddy:2022knd}, to accurately account for Sommerfeld enhancement~\cite{Blum:2016nrz}, re-summation of Sudakov double logarithms~\cite{Baumgart:2017nsr}, and the inclusion of continuum photons near the endpoint of the spectrum~\cite{Beneke:2022eci}.      

WIMP DM connected to supersymmetry, and natural solutions to the hierarchy problem more generally, remains a promising DM framework; for instance Twin Higgs models also contain a variety of well-motivated DM candidates \cite{Asadi:2022njl}.  However, the past decade has seen enormous progress, led by theoretical efforts, in constructing models of particle DM that go beyond the WIMP paradigm.  This explosion of DM model-building has substantially developed ideas for how DM may be produced in the early universe as well as how it may be related to other unsolved issues in particle physics.  This theoretical effort has likewise substantially broadened the range of DM signatures of interest in indirect detection, direct detection, and collider experiments.
Further discussion of specific particle DM models can be found in the contributed white papers \cite{Asadi:2022njl, Ghosh:2022zef, Agrawal:2022rqd, Carney:2022gse, Dienes:2022zbh}; 
  here and in the next two sections we discuss implications for the discovery prospects of DM in both terrestrial and astrophysical experiments.

\subsection{Indirect Detection in Cosmic Rays}

The indirect detection of DM annihilation or decay in the local universe through its imprint on cosmic rays is a key element of the program to detect and identify DM\null.   
In recent years, indirect detection analyses have made major progress in exploring the parameter space where signals are predicted in the most straightforward models, such as WIMP DM and sterile neutrino DM\null.  These two well-motivated models have historically provided much of the motivation for indirect detection searches. 
Theorists have been key parts of this effort, and have developed new analytical strategies and analysis tools for these searches \cite{Boddy:2022knd,Leane:2022bfm}.
Progress in  experimental tests of these two landmark signatures has been accompanied by a substantial broadening of astrophysical searches for the imprint of DM in cosmic rays.  
Recent developments in DM model-building have  expanded the range of DM annihilation and decay signatures that are of interest to indirect detection, providing models with novel spectral signatures, signatures that populate new kinematic areas, signatures in novel final states, and signatures with unusual spatial distributions \cite{Asadi:2022njl, Dienes:2022zbh, Ghosh:2022zef, Carney:2022gse}.

High-energy photons in the $X$-ray and gamma-ray bands
are among the most sensitive cosmic ray channels for detecting DM annihilation and decay in the low-redshift universe.  As experimental searches push into new regimes at both the high and low ends of this energy range, new theoretical work is needed in many directions in order to ensure the full success of the experimental program.  

For one illustrative example, upcoming observatories such as the
CTA, the Large High Altitude Air Shower Observatory (LHAASO), and the Southern Wide-field Gamma-ray Observatory (SWGO) \cite{Aramaki:2022zpw} will deliver substantial new experimental sensitivity to very high energy photons.  However,  
accurately characterizing the signals of DM annihilation in the TeV regime where these experiments have sensitivity has required a broad theoretical program to establish the effects of electroweak interactions in both the size of the overall DM annihilation cross-section as well as the resulting photon and cosmic ray spectra~\cite{Bauer:2014ula,Ovanesyan:2014fwa}.
In addition, understanding the kinds of DM that experiments such as CTA, LHAASO, SWGO, and IceCube can discover relies on the active theoretical development of
models of ultra-heavy DM, which generally feature substantially different cosmologies and/or production mechanisms than models of lighter DM~\cite{Carney:2022gse}.  

For lower energy photons, e-Astrogram and AMEGO in the  MeV range and ATHENA in the X-ray band are expected to substantially increase the sensitivity to DM signatures of keV-GeV DM\null.  These missions will benefit from the large number of novel analysis techniques for DM annihilation and decay that have been developed over the past decade.
Examples of notable theoretical development over the past decade that will influence analyses of future data include \cite{Boddy:2022knd}: (i) identifying new targets for DM annihilation and decay searches, ranging from newly-discovered Milky Way satellites to possible DM-only halos to the column density of the Milky Way itself; (ii) developing novel analysis techniques such as the cross-correlation of signals in different data sets and machine-learning based techniques; and (iii) improvements in the modeling of astrophysical backgrounds.

As a particularly well-motivated case, consider the  Galactic Center gamma-ray excess (GCE) observed in Fermi data.  %.  
Understanding the origin of this novel gamma-ray emission is one of the most urgent tasks facing indirect detection in the gamma-ray band. Work by theorists has been a main driver in the development of novel analysis techniques and alternate searches to probe the possible origin of the GCE, whether DM or otherwise \cite{Leane:2022bfm}.  
Substantial theory work remains key to identifying the origins of the GCE.  There are many important directions where theoretical work is needed to improve our understanding of both astrophysical and DM contributions; a common thread among them is the use of intensive modeling to correlate input from multiple sources of information (charged cosmic rays, photons in multiple wavelengths, hydrodynamic simulations, measurements of stellar kinematics, stellar population simulations, and more)  \cite{Leane:2022bfm}.  
 
The GCE provides an excellent example of a case where innovative new methods from the theoretical community have played an important role in identifying and characterizing the excess, and also highlights the importance of a close connection between the theoretical and observational communities.  Attaining the best discovery prospects in indirect detection in general will rely on both the continued creativity of the theoretical community and the synergistic dialogue between theorists and observers.

\vspace{.5 cm}

\section{Axions}
\label{sec:axions}

The QCD axion is one of the most promising candidates for new physics.  The chance for discovering this particle, if it exists, over the next decade is high, thanks in large part to recent theoretical breakthroughs. Axions have deep connections to a range of topics in theoretical physics including DM, cosmology, and string theory, as we elaborate on below.  Axions are light, bosonic particles that are part of a broader class of possible new physics candidates that includes dark photons and scalars, which could also be DM candidates. In this report we focus on axions, which provide a well-established and especially well-motivated example of these wave-like DM candidates.   
However, scalars and dark photons, especially in the context of wave-like DM, have also seen enormous theoretical progress over the past decade and also have compelling ultraviolet and phenomenological motivations, as summarized in the Snowmass2021 white paper~\cite{Antypas:2022asj} specifically on these particles. 

\subsection{Axion Models}

The QCD axion was originally introduced as a theoretical construct to address the strong-CP problem. QCD coupled to massive quarks has a CP-violating angle $\bar \theta$ that generates a neutron electric dipole moment (EDM): $d_n \sim 10^{-16} \, \, {\rm e} \cdot {\rm cm}$ (see~\cite{Kim:2008hd,Hook:2018dlk} for reviews).  However, direct searches for the neutron EDM have yet to yield a non-zero value, constraining $|d_n| \lesssim 3 \times 10^{-26}$ ${\rm e} \cdot {\rm cm}$~\cite{Baker:2006ts}, which implies that $|{\bar \theta}| \lesssim 10^{-10}$.  Given the non-zero CP violating phase in the Cabibbo-Kobayashi-Maskawa (CKM) matrix and P violation in the electroweak sector, CP and P are not restored as symmetries of the SM if ${\bar \theta} = 0$, making symmetry-based solutions to the strong-CP problem difficult.  With that said, a number of symmetry-based solutions to the strong-CP problem have been proposed, based upon spontaneously broken CP or P symmetry~\cite{Choi:1992xp,Dine:1992ya,Babu:1988mw,Babu:1989rb,Barr:1991qx,Nelson:1983zb,Barr:1984qx}.  A crucial element of spontaneously broken P models is the existence of an extended electroweak sector with {e.g.} $SU(2)_R$ and new strongly-interacting fermions charged under the right-handed $SU(2)$ sector.  Direct searches for new physics associated with this extended sector are promising avenues for new physics searches at the LHC and future colliders~\cite{DAgnolo:2015uqq,Craig:2020bnv}.

In contrast to the symmetry-based solutions, the axion is a dynamical solution to the strong-CP problem.  The axion is postulated as a pseudo-Goldstone boson of a symmetry, the Peccei-Quinn (PQ) symmetry, that is broken at a high scale $f_a$; the axion $a$ would be massless but for its coupling to QCD ${\mathcal L} \sim a G \tilde G / f_a$, where $G$ is the QCD field strength, $\tilde G$ is its dual, and color and spacetime indices have been suppressed~\cite{Peccei:1977hh,Peccei:1977ur,Weinberg:1977ma,Wilczek:1977pj}.  Below the QCD confinement scale QCD instantons generate a potential for the axion $V(a) \sim \Lambda_{\rm QCD}^4 (a / f_a + \bar \theta)^2$, with $\Lambda_{\rm QCD}$ the QCD confinement scale and with interaction terms not shown; when the axion minimizes its potential it dynamically removes the neutron EDM, which is proportional to the same combination $d_n \propto (a/f_a + \bar \theta)$ that enters into the potential.  In addition  to acquiring a small mass $m_a \approx 0.57 \, \, {\rm neV} (10^{16} \, \, {\rm GeV} / f_a)$, the axion acquires interactions with electromagnetism of the form ${\mathcal L} \supset g_{a\gamma\gamma} a F \tilde F / 4$, with $g_{a\gamma\gamma} = C_{a\gamma\gamma} \alpha_{\rm EM}/(2 \pi f_a)$ the axion-photon coupling, with $C_{a\gamma\gamma}$ a dimensionless and model-dependent number, and $F$ the quantum electrodynamics (QED) field strength.  The axion also generically has dimension-5 derivative couplings to fermions.
See~\cite{DiLuzio:2020wdo} for a modern review of axion models that make specific predictions for the axion-photon and axion-matter couplings.

Apart from the strong-CP problem axions are motivated for purely theoretical reasons because they arise generically in the context of string theory constructions (see, {e.g.},~\cite{Svrcek:2006yi}).  The recent work on axions illustrates the synergy between formal theoretical work, applied phenomenological research, and experimental searches for new physics.  For example, 
the last decade has seen significant progress in understanding the properties of axions that emerge in the string landscape, where quantum gravity effects are calculable.  Ref.~\cite{Demirtas:2021gsq} computed the PQ-violating contributions to the QCD axion potential in an ensemble of $\sim$100,000 type IIB string theory compactifications and found that in 99.7\% of those compactifications the quantum gravity contributions to the neutron EDM ({i.e.}, the extent to which the PQ symmetry is broken) were below the current experimental precision.  This result reflects the trend observed in multiple recent works based off of {e.g.}~type IIB string theory,  F-theory, heterotic string theory, and M-theory;
high-quality axions arise generically in these constructions.  On the other hand, the string theory constructions have led to the idea of the axiverse~\cite{Arvanitaki:2009fg},
which is the observation that string theory compactifications typically do not just produce one axion but rather a large number $N \sim 10'{\rm s} - 100'{\rm s}$ of axion-like particles, due to the number of higher-dimensional gauge fields and the complexity of the compacified manifold.   Axion-like particles are commonly referred to as ALPs (though we will refer to them as axions, to be differentiated from the QCD axion).  Axions may also be heavy, and in particular the $\sim$keV - GeV mass range has been shown recently to lead to novel astrophysical and terrestrial signatures (see, {\it e.g.},~\cite{Baryakhtar:2022obg,Essig:2022yzw}).

\subsection{Axion Dark Matter}

Shortly after the discovery of the QCD axion as a theoretical construct for solving the strong-CP problem it was realized that cosmologically-produced, free axions could explain the observed DM density~\cite{Preskill:1982cy,Abbott:1982af,Dine:1982ah}.  
The production mechanism for QCD axion DM depends crucially on the ordering of inflation relative to PQ symmetry breaking (see~\cite{Essig:2022yzw,Baryakhtar:2022obg,Blinov:2022tfy,Adams:2022pbo,Boddy:2022knd} for discussions of axion DM in the context of Snowmass2021).  If the PQ symmetry is broken before or during inflation and is not restored after inflation, then the DM abundance is determined by the misalignment mechanism, which states that the cosmic axion field $a(t)$ evolves, for generic initial conditions, according to the homogeneous and linearized equation $\ddot a + 3 H \dot a + m_a^2(T) a =0 $, with derivatives with respect to cosmic time $t$, $H$ the Hubble parameter, and $m_a^2(T)$ the temperature-dependent axion mass. 

To compute the abundance of axion DM one must track the temperature-dependent axion mass, which vanishes at temperatures well above the QCD phase transition and which asymptotes to its zero-temperature value below the QCD phase transition.  In particular, we may relate the temperature-dependent mass to the temperature-dependent QCD topological susceptibility $\chi_{\rm top}(T)$ through the relation $m_a^2(T) = \chi_{\rm top}(T) / f_a^2$.  For temperatures $T \gg T_c$, with $T_c \sim 170$\,MeV the QCD critical temperature, the topological susceptibility may be computed in the dilute instanton gas approximation.  However, lower temperatures require lattice QCD simulations.  The lattice QCD results thus play a crucial role in determining the axion DM abundance, with state-of-the-art predictions emerging over the past decade from dedicated lattice efforts~\cite{diCortona:2015ldu,Borsanyi:2016ksw,Ballesteros:2016xej}.

On the other hand, it is possible that the PQ symmetry is restored after inflation and spontaneously broken in the subsequent thermal cooling of the universe.  In this case the computation of the DM abundance is more complicated because the initial misalignment angle varies from Hubble patch to Hubble patch at PQ symmetry breaking, and yet these patches come into causal contact with each other during the subsequent evolution.  Topological defects known as axion strings arise and play an important role in determining the final DM abundance, as they radiate axions during their evolution.  Moreover, leading up to the QCD phase transition domain walls develop between the strings, which have tension and cause the string network to collapse, further producing axions that contribute to the DM abundance.  
Attempts have been made for decades to calculate the DM abundance in this cosmological scenario, with predictions ranging from $m_a \sim 10$ $\mu$eV to $m_a \gg 500$ $\mu$eV~\cite{Klaer:2017ond,Gorghetto:2018myk,Buschmann:2019icd,Gorghetto:2020qws}.
The past decade, however, has seen a revolution in the complexity and accuracy of axion cosmology simulations, aided in large part by advances in high-performance computing.  
Modern axion cosmology simulations take full advantage of cutting-edge supercomputing facilities, and these computations will improve in accuracy further in the near future by leveraging computational and technological advances.    
As of 2022 the leading prediction for the axion mass is $m_a \sim 100$ $\mu$eV~\cite{Buschmann:2021sdq}, but this result remains contentious~\cite{Gorghetto:2020qws,Martins:2018dqg,Hindmarsh:2021vih}.
Future effort is needed to further refine the mass prediction.  This theoretical work is crucial for the experimental program because it informs which mass range experiments should target; for example, the results to date point to axion masses above those probed by most resonant cavity experiments.  This example illustrates the importance of theoretical work in directly informing which experiments should be focused on to have the best chances of discovering DM\null.  See the Snowmass2021 white paper~\cite{Asadi:2022njl} for a more extensive discussion of axion DM production mechanisms.

\subsubsection{Fuzzy Dark Matter}

Axion-like particles, which do not couple to QCD but have small bare masses or small masses from confining dark sectors, may also make up a sizeable fraction of the DM\null.  The axion-like particles could achieve their relic abundance either through the misalignment mechanism, like the QCD axion, from topological defects like axion strings, or through more exotic production mechanisms~\cite{Co:2019wyp}.  Indeed, the misalignment production mechanism
predicts that the correct DM abundance is achieved for axion masses $m_a \sim 10^{-22}$ eV with GUT-scale decay constants; such DM candidates are referred to as fuzzy DM (see, {e.g.},~\cite{Hui:2016ltb}).  Fuzzy DM has received a boom of interest over the past decade because of possible connections to string theory, which as discussed may give rise to a large number of axion-like particles and non-perturbative mechanisms for generating exponentially-suppressed masses at the right mass scale, and because of the fact that fuzzy DM modifies structure on astrophysical scales (see~\cite{Hui:2021tkt} for a recent review and~\cite{Bechtol:2022koa} for a discussion in the context of Snowmass2021).  In particular, fuzzy DM suppresses the number of DM halos at masses less than $\sim$$10^{10}(m_a / 10^{-22} \, \, {\rm eV})^{-3/2}$ $M_\odot$, while the central cores of fuzzy DM halos may be solitonic~\cite{Hui:2021tkt}.  Fuzzy DM may also reduce the dynamical friction that objects feel in DM halos~\cite{Hui:2016ltb}.  More generally, over the past decade fuzzy DM received a wave of interest in part to explain a number of apparent failings of the standard cosmological model on small astrophysical scales, including the lack of low-mass subhalos of the Milky Way and the ``too-big-to-fail" problem associated with the lack of high-luminosity satellite galaxies.

On the other hand, fuzzy DM has become increasingly constrained in the latter half of the past decade, thanks to theoretical efforts in understanding the astrophysical implications of fuzzy DM, cosmological and galactic-scale simulations incorporating fuzzy DM, and new data (see~\cite{Banerjee:2022qcb,Boddy:2022knd} for discussions in the context of Snowmass2021).  The DM mass is now constrained to be $m_a \gtrsim 2 \times 10^{-20}$ eV from analyses looking for suppressed structure on small scales in the Lyman-alpha forest~\cite{Rogers:2020ltq}, while direct searches for suppressed power in the halo-mass function constrain $m_a \gtrsim 2 \times 10^{-21}$ eV~\cite{Schutz:2020jox}.  Recently, masses less than $\sim$$10^{-20}$ eV have been shown to be disfavored by rotation curve data~\cite{Bar:2021kti}.  
These recent results, while unfortunately null, represent a triumph of the theoretical work that has been performed over the past decade.
(See the Snowmass2021 white papers~\cite{Alvarez:2022rbk} and~\cite{Bechtol:2022koa} for a more thorough discussion of fuzzy DM\null.)

\subsection{Axion Indirect Detection}

It has long been understood that axions could leave detectable astrophysical signatures by modifying stellar cooling.  However, in the past decade a number of novel astrophysical probes of axions and axion DM have been developed, in part because of input from string theory motivating axion-like particles and in part from considering a broader parameter space for the QCD axion.  At the same time, the stellar cooling probes have been refined, such that at present the strongest constraints on the QCD axion and axion-like particles, apart from a narrow mass range probed by the ADMX experiment, arise from astrophysical probes (see~\cite{Boddy:2022knd,Baryakhtar:2022obg} for a discussion in the context of Snowmass2021).  Theorists have been crucial in developing the ideas behind these searches and implementing them with astrophysical data.

For example, the axion-nucleon sensitivity has recently been improved by neutron star (NS) cooling modeling efforts, 
constraining the QCD axion mass to be less than $\sim 10$--$30$ meV, depending on the ultra-violate (UV) completion~\cite{Buschmann:2021juv}.  Recent modeling efforts for axion production in the context of SN 1987a produce comparably strong results~\cite{Carenza:2019pxu}.  In total, thanks to a variety of creative stellar cooling probes that have been developed and implemented over the past decade, QCD axions with masses above $\sim$10 meV are strongly disfavored, though intriguingly stellar cooling anomalies do exist, for example in the context of WD cooling, which could be explained by a lower-mass QCD axion~\cite{Giannotti:2017hny,Hoof:2018ieb,DiLuzio:2021ysg}.   

The realization that axion-like particles with coupling strengths slightly below current constraints are well motivated by high-scale modeling efforts, such as string theory axion constructions~\cite{Halverson:2019cmy}, has helped stimulate a wave of brand new ideas for probing axions using astrophysical data over the past decade.  For example, it was shown that axion DM may convert into narrow radio lines around NS magnetospheres, drawing radio telescopes into the search for axion DM and enriching the science case for upcoming missions such as the square kilometer array (SKA). Similarly, in seminal papers~\cite{Arvanitaki:2009fg,Arvanitaki:2010sy} at the beginning of the past decade it was shown that black hole superradiance could probe ultralight axions motivated by string theory and the axiverse, catalyzing the creation of an entire subfield now spanning the theoretical and observational communities, including the gravitational wave community, looking for effects of axion-induced black hole superradiance. 
See the Snowmass2021 white papers~\cite{Boddy:2022knd,Baryakhtar:2022obg,Brito:2022lmd} for more thorough discussions.

A key point behind theoretical work in astrophysical probes of new physics is the ability of research to connect formal aspects of theoretical physics with cutting-edge data through astrophysical modeling.  
For example, black hole superradiance arose from  curiosity-driven investigations of black holes ({e.g.},~\cite{PhysRev.93.99,1971NPhS..229..177P,1971JETPL..14..180Z,1973JETP...37...28S}), which evolved to the realization that superradiance could play a relevant role in astrophysics if ultralight axions exist in nature ({e.g.},~\cite{Arvanitaki:2009fg}).  These initial works were followed by hundreds of dedicated phenomenological studies, including projected sensitivities and search strategies using near-term instruments, investigations of the broader particle parameter space to {e.g.} include spin-1 particles and the effects of self-interactions, and evolution in more complicated environments such as binary systems~\cite{Baryakhtar:2022obg}.  Now, specific, simulation-based predictions for current and near-term gravitational wave detectors provide benchmarks for the gravitational wave science case.

\subsection{Axion Direct Detection}

The last decade has seen a paradigm shift in the way that axion DM is searched for in the laboratory (see~\cite{Adams:2022pbo,Essig:2022yzw} for Snowmass2021 summaries).  Before $\sim$2013 terrestrial axion experiments were essentially of three types: (i) light shining through walls experiments; (ii) axion helioscopes looking for axions produced in the Sun; and (iii) axion resonant microwave cavity haloscopes looking for axion DM\null.  These latter two experimental pathways were proposed originally in a seminal theoretical paper by Sikivie from the early 1980's~\cite{Sikivie:1983ip}, with the former proposal originating later in the 1980's~\cite{VanBibber:1987rq}.  Sikivie's original theoretical work in particular on axion direct detection laid the groundwork for the majority of the following nearly four decades of experimental effort to detect axions in the laboratory.  Until recently there were two main experimental collaborations for axion laboratory detection: the CERN Axion Solar Telescope (CAST) axion haloscope experiment~\cite{CAST:2015qbl} and  the Axion Dark Matter Experiment (ADMX) G2~\cite{ADMX:2018gho,ADMX:2019uok,ADMX:2021nhd}, which is currently one of the three US flagship DM experiments.  

ADMX is a resonant microwave cavity experiment, which works by applying a strong, static magnetic field inside a high-quality factor microwave cavity.  Axion-to-photon conversion takes place in the cavity, and if the cavity resonant frequency is tuned to the axion mass the resulting signal may be further amplified.  A tuning rod is used to adjust the cavity resonant frequency and thus scan over different axion masses
(see~\cite{Adams:2022pbo} for additional planned microwave cavity efforts).  However, despite the success of the microwave cavity haloscope efforts, there are clear limitations with this approach when going beyond $\sim$1-100 $\mu$eV in axion mass.  First, the cavity size should be roughly matched to the axion Compton wavelength, which is $\lambda \approx 25 \, \, {\rm cm}\,\, (5 \,\, \mu{\rm eV} / m_a )$; axion masses $m_a \ll 10^{-6}$ eV would thus require unreasonably large cavities.  On the other hand, the signal power is proportional to the cavity volume, which scales like $\lambda^3$; at higher masses it thus becomes difficult to achieve the necessary signal-to-noise ratio, amongst additional experimental issues~\cite{Adams:2022pbo}.

Thanks to pioneering theoretical effort, followed by bold and innovative small-scale experimental programs, there now exist proposals for probing nearly the entire currently-allowable QCD axion mass range.  An important theoretical observation at the beginning of the last decade was that the axion, while it solves the strong-CP problem by removing the time-averaged neutron EDM, leaves a time-varying, residual EDM when the axion is DM~\cite{Graham:2011qk,Graham:2013gfa}.  Since the EDM is proportional to the axion field, the oscillation frequency is simply the axion mass.   This led to the proposal for the CASPEr experiment~\cite{Budker:2013hfa}, which aims to detect the oscillating axion-induced EDM using a nuclear magnetic resonance based experiment.  That proposal is now an experimental collaboration pursuing GUT-to-Planck scale QCD axion DM\null.   
A similarly ground-breaking discovery came from the observation that by thinking more broadly about the modifications to Maxwell's equations in the presence of axion DM, experimental setups could in principle be constructed that would be sensitive to GUT scale axion DM using the axion-photon coupling~\cite{Sikivie:2013laa,Kahn:2016aff}.  In the limit where the axion Compton wavelength is much larger than the size of the experiment, axion DM modifies electromagnetism so as to induce effective currents that follow laboratory magnetic field lines and oscillate at the frequency of the axion mass.  In the setup in~\cite{Kahn:2016aff}, for example, a toroidal magnet in the laboratory would induce an effective circular, oscillating current density in the presence of axion DM, which creates a real, oscillating magnetic flux through the center of the toroid; the proposal is to detect this oscillating flux using a pickup loop and SQUID circuit.  Moreover, the induced signal is narrow in frequency space, meaning that it may be resonantly enhanced using an LC circuit~\cite{Sikivie:2013laa,Chaudhuri:2014dla,Kahn:2016aff}.  
The theoretical proposal in~\cite{Kahn:2016aff} became the basis of the ABRACADABRA 10-cm collaboration~\cite{Ouellet:2019tlz}.
The ABRACADABRA collaboration has joined with the DM Radio collaboration to propose multiple experiments over the next decade to probe QCD axion DM with decay constants near and below the GUT scale. See~\cite{Adams:2022pbo,Essig:2022yzw} for discussions in the context of Snowmass2021 of additional novel theoretical ideas for probing axion DM, and low mass DM generally, many of which have led to the creation of dedicated experiments.

\subsection{Axions in Cosmology}

Axions are a striking example of how the history of the universe is interconnected with the laws of nature.  They could be essential to cosmic acceleration, at early times via inflation, as a possible source of dark energy today, or at any time in between (the latter being motivated by attempts to address the Hubble tension).  Furthermore, both hot and cold components of the axion field are important for the evolution of the universe starting from the hot big bang through structure formation.

Axions have played a dominant role in our inflationary model building both in string theory and field theory, dating back to the advent of natural inflation~\cite{Freese:1990rb}.  As the potential is only generated non-perturbatively, axion models of inflation are typically more robust to quantum corrections.  Continued exploration of these models has consistently unearthed new phenomena, including monodromy~\cite{Silverstein:2008sg,McAllister:2008hb} and novel sources of particle production.  For high scale inflation, the axion decay constant that would be needed exceeds that Planck scale, motivating the exploration of whether such models can arise from a consistent $\Lambda$CDM completion~\cite{Draper:2022pvk}.  Axions may also be light spectators during inflation.  Even without playing an active role during inflation, these fields will still be excited from the vacuum by the expansion of the universe and correspond to isocurvature modes during inflation.

After inflation, the universe reheats and gives rise to the hot big bang.  The SM particles remain in thermal equilibrium as the universe cools, up until neutrino decoupling begins at temperatures of a few MeV\null.  Axions can be thermalized at any point during this epoch, depending on the strength of the coupling to the SM\null.  If thermalized, these hot axions will contribute to $\Neff$ as measured through the primordial abundance of elements, the CMB, and the LSS of the Universe~\cite{Dvorkin:2022jyg}.  In the absence of a large number of new massive degrees of freedom or departure from equilibrium, each axion will contribute $\Delta \Neff \geq 0.027$, even if the axion decouples near the reheating temperature.  This is an exciting target that is driving the design of future cosmic surveys.

In contrast, if the PQ symmetry is not restored after inflation, any isocurvature fluctuations produced during inflation will survive through today.  Constraints on isocurvature around the era of recombination are strongly probed by the CMB\null.  In addition, these isocurvature modes will also impact the Baryon Acoustic Oscillations (BAO) and are a potential target for future large scale structure surveys.

\section{Light Dark Matter and Dark Sectors}

The past decade marked a turning point in the approach to understanding the microscopic nature of DM\null.  As already stressed, prior to the LHC era, WIMP DM was the standard paradigm that drove theoretical and experimental efforts.  However, the lack of data supporting WIMP DM (or other new physics at the electroweak scale, such as supersymmetry) has caused the community to question whether WIMP DM is realized in nature. Generally, this question has been the driving force behind progress in the search for DM over the past decade.  On the one hand, as discussed in the previous sections, a significant amount of recent theoretical work has been focused on better probing long-motivated DM models such as the QCD axion and currently (or previously) unprobed regions of WIMP DM parameter space, such as the supersymmetric winos and higgsinos.  On the other hand, there has been an increasing push to develop bottom-up DM models, beyond the WIMP, and search strategies that are signature-driven and built upon EFT principles, instead of being motivated by ultraviolet considerations.  

A key area where this new approach to DM model building and detection has seen rapid progress is that of sub-GeV dark sectors~\cite{Essig:2013lka,Alexander:2016aln,Battaglieri:2017aum,Agrawal:2021dbo}.  The basic idea behind such constructions is to postulate additional, light BSM particles that are singlets under the SM gauge symmetry and may or may not have self-interactions, which may be relatively strong.  New particles of nature with masses at or below the GeV scale must necessarily interact very feebly with the SM in order to be consistent with tests of the SM\null. If DM resides in the dark sector it can talk to the SM through a number of feebly-coupled portals.  This is contrasted with {e.g.}~the WIMP paradigm where the DM is charged under the electroweak force; in the dark sector paradigm, the DM is neutral under the SM\null.  Dark sector constructions often contain light mediators that can be directly probed by experiments.  This differs from axion models, where the mediating particles are often extremely heavy, with masses as high as the GUT scale.

One canonical category of dark sector models is that where the mediator is a dark photon that kinetically mixes with the ordinary photon~\cite{Holdom:1985ag}.  In a minimal construction the dark photon is a portal to a dark fermion or dark scalar that is the DM candidate (see, {e.g.},~\cite{Alexander:2016aln}).  The DM abundance may be achieved through standard thermal freeze-out by off-shell dark photon production, with a sub-GeV mass possible, depending on the dark photon mass and coupling constant.  However, powerful constraints from indirect detection, such as the CMB~\cite{Planck:2018vyg,Slatyer:2015jla}, require that the DM be heavier than tens of GeV if it annihilates through $s$-wave processes (this is the case for the above example of a dark fermion annihilating through an off-shell dark photon).  These constraints are circumvented if DM annihilates though $p$-wave processes (such as a dark scalar in the above example or a Majorana fermion), or if DM is produced through another mechanism, as discussed below.  

An intriguing and generic possibility, however, is that where the portal interaction is too weak to ever cause the DM to come into thermal equilibrium with the SM\null.  
In this scenario a relic DM population can be produced through the ``freeze-in" mechanism, where DM is produced slowly by interactions, such as decays or scattering, that are too weak to thermalize the DM~\cite{Hall:2009bx,Essig:2011nj,Chu:2011be}.   In the limit of a light mediator, the minimal dark photon portal model can give a precise prediction for the dark photon couplings needed to get the correct DM abundance from freeze-in, given a DM mass.  This, in turn, leads to a precise prediction for DM scattering rates at direct detection experiments~\cite{Essig:2011nj,Alexander:2016aln,Fernandez:2021iti}.  The minimal freeze-in scenario has become a benchmark for experimental efforts pushing to lower DM masses, as we discuss shortly.  DM produced by freeze-in is an example of a Feebly Interacting Massive Particle (FIMP).  In FIMP models~\cite{Bernal:2017kxu}, the DM interactions are too weak to lead to thermalization, unlike the WIMP scenario, where the DM-SM interactions are strong enough to thermalize the DM\null.  Other example of FIMPs include SuperWIMPs (also known as freeze-out and decay)~\cite{Feng:2003xh} and the production of sterile neutrinos from oscillations~\cite{Dodelson:1993je}.

DM masses at or below the GeV scale are now known to arise in a variety of novel dark sector scenarios.  Examples include asymmetric DM models that relate DM and baryon asymmetries~\cite{Kaplan:2009ag,Falkowski:2011xh}, Strongly Interacting Massive Particles (SIMPs), where the DM abundance is determined by the freeze-out of $3 \to 2$ processes~\cite{Hochberg:2014dra,Hochberg:2014kqa}, semi-annihilations where two DM particles annihilate to one DM particle and another state~\cite{DEramo:2010keq}, forbidden channels where DM annihilates into heavier states~\cite{Griest:1990kh,DAgnolo:2015ujb}, co-annihilations where multiple states participate in the annihilations~\cite{Griest:1990kh,DAgnolo:2018wcn}, and coscattering where DM is produced by scattering instead of annihilations~\cite{DAgnolo:2017dbv}.

\subsection{Light Dark Matter Direct Detection}

Prior to the early 2010s, the experimental effort for seeking the direct detection of DM was singularly focused on elastic or inelastic scattering off of nuclei.
However, at the beginning of the past decade, motivated by sub-GeV dark sector modeling efforts, a range of theoretical proposals was put forth for detecting novel inelastic DM interactions with electrons in materials (see, {e.g.},~\cite{Bernabei:2007gr,Essig:2011nj,Graham:2012su,Lee:2015qva}).  The advantage of scattering off of electrons for light DM is clear: since the electron has less mass, it acquires a larger recoil energy relative to nuclei during scattering processes.  Since electrons tend to be bound, DM-electron processes tend to be inelastic ({e.g.,} ionization).  

Theoretical work was crucial in producing the first constraints on sub-GeV DM, down to a few MeV, by searching for ionization signatures from DM-electron scattering in XENON10 data~\cite{Essig:2012yx}.  It was also proposed that semi-conductor targets might be even better suited for light-DM searches, since the DM would in principle just need enough energy to push an electron over the band-gap in order to produce a detectable signature, though calculating the scattering rates is non-trivial and connects with cutting-edge topics in condensed matter physics~\cite{Essig:2011nj,Graham:2012su,Lee:2015qva,Essig:2015cda}.

The theoretical work showing the promise of semiconductor targets for light DM motivated an experimental research program to try to detect one or few electron events, which would make the detectors sensitive to DM depositing just enough energy to push an electron above the band gap (corresponding to {e.g.}~500 keV DM for scattering and $\sim$1 eV DM if absorbed, for a Si target).  The detection of single electron scattering events is now possible, as demonstrated by the SuperCDMS and SENSEI collaborations~\cite{Tiffenberg:2017aac,Romani:2017iwi}.  On the experimental side, reaching the thresholds pointed out in the theoretical works has brought about the successful demonstration of multiple types of novel quantum sensors, such as Skipper-CCDs and high-voltage charge amplification transition edge sensors.  The theoretical and experimental communities joined forces to form the SENSEI collaboration, which has already produced first results~\cite{Crisler:2018gci,SENSEI:2019ibb,SENSEI:2020dpa} and which  aims to detect sub-GeV DM at the freeze-in level, and SuperCDMS, DAMIC, and EDELWEISS have also achieved comparable sensitivity to DM-electron scattering events in solid-state detectors~\cite{SuperCDMS:2018mne,SuperCDMS:2020ymb,DAMIC:2019dcn,EDELWEISS:2020fxc}.  Larger solid-state detectors, building off of the original theory proposals and successful experimental demonstrations, are now being planned and promise to detect or exclude the freeze-in DM benchmarks~\cite{Castello-Mor:2020jhd,Aguilar-Arevalo:2022kqd}.  Furthermore, there continue to be fruitful theoretical proposals, often followed through by small-scale experimental demonstrations, for pushing the low-threshold frontier, such as searching for DM scattering in scintillating crystals~\cite{Derenzo:2016fse}, aromatic organic targets~\cite{Blanco:2019lrf}, two-dimensional targets with directional information~\cite{Hochberg:2016ntt}, three-dimensional Dirac materials~\cite{Hochberg:2017wce}, and superconducting nanowires~\cite{Hochberg:2019cyy}, amongst other possibilities (see~\cite{Essig:2022yzw} for a more extensive discussion for Snowmass2021).  On the other hand, the low-threshold detectors will only prove useful if the backgrounds are minimized with well-understood rates.  Recently, theoretical work has contributed significantly in this direction (see, {e.g.},~\cite{Du:2020ldo}). Future theoretical work on understanding the background rates in novel detectors will be crucial for the success of the low-threshold experimental program. 

DM with mass below $\sim$keV should be bosonic in order to satisfy the Tremaine-Gunn bound~\cite{1979PhRvL..42..407T}, which states that fermions with too low masses cannot make up the DM because the largest-density DM objects, such as dwarf galaxies, would require multiple particles to occupy single quantum states, which is not allowed for fermions.  However, there are many well motivated sub-keV bosonic DM models, such as axion and dark photon models.  In the limit where these particles have masses much less than an eV, it is appropriate to think of the DM as classical waves, while if the DM has mass well above the eV scale then the particle language is more appropriate.  However, it has only become clear how to approach the detection of $\sim$eV scale bosonic DM candidates recently, when it was pointed out that they could be absorbed in photonic materials~\cite{Baryakhtar:2018doz}.  These so-called ``optical haloscopes" operate using alternating dielectric layers to enhance the axion-photon conversion probability at optical frequencies (see the MADMAX experiment~\cite{MADMAX:2019pub} for a similar idea applied to lower frequencies). 
The theoretical proposal in~\cite{Baryakhtar:2018doz} has already led to the creation of two experimental collaborations searching for eV-scale axions and dark photons: LAMPOST~\cite{Chiles:2021gxk} and MuDHI~\cite{Manenti:2021whp}.   

\subsection{Accelerator Searches for Dark Sectors}

The realization that DM may reside in a dark sector feebly coupled to the SM has come with the understanding that in many dark sector models the most promising detection strategy is to search for the mediator itself in accelerator-based experiments.  The search for feebly-coupled mediators at collider experiments has emerged as a major new discovery program and is distinct from the traditional approach to new physics searches at colliders, which tends to focus on  probing new, relatively strongly-coupled states with masses closer to the energy frontier.  In fact, light dark sector models are often most strongly constrained by their accelerator signatures, as elaborated in a number of recent theoretical works (see, {e.g.},~\cite{Fabbrichesi:2020wbt} and references therein).  Theorists have played crucial roles not only in introducing the theoretical concept of dark sectors and developing related concepts, like the phenomenology of long-lived particles, but also in proposing concrete experiments to search for such new physics, many of which have already taken data or are in the construction phases.  Below, we highlight a few examples where theorist-driven proposals have matured into accelerator-based experimental collaborations.  (See~\cite{Essig:2022yzw} for a more extensive discussion in the context of Snowmass2021.)    

In a seminal theoretical paper around a decade ago, it was pointed out that  results from a number of existing beam dump experiments could be reinterpreted to set strong constraints on massive dark photons, like those found in many dark sector scenarios, and that new dedicated accelerator experiments, using existing technology, could significantly expand the reach to dark photon and dark sector scenarios~\cite{Bjorken:2009mm}. This led to a number of theoretical proposals for dedicated or auxiliary dark sector accelerator-based searches over the past decade, many of which have already turned into actual experiments or are funded and in construction phases.  For example, the APEX experiment at Jefferson Lab~\cite{Essig:2010xa}, which is an electron fixed-target experiment looking for dark photon production following by decays to $e^+e^-$, was a direct product of the proposals in~\cite{Bjorken:2009mm}, as was the dark photon search at the Mainz Microtron~\cite{A1:2011yso} and the Heavy Proton Search Experiment~\cite{Battaglieri:2014hga}.  

Dark sector DM may also be generated directly in collider experiments.  Theoretical work~\cite{deNiverville:2011it} pointed out that this strategy could also be applied to accelerator-based neutrino experiments, such as LSND and MiniBooNE, which went on to become part of the neutrino experiments' science program~\cite{MiniBooNE:2017nqe,MiniBooNEDM:2018cxm}.  It was also realized that electron beam dumps, followed by down-stream detectors, could also be used to search for dark-sector DM candidates~\cite{Izaguirre:2013uxa,BDX:2014pkr}.

Theorists have similarly pointed out that LHC experiments can operate symbiotically with down-stream or nearby auxiliary detectors designed to search for long-lived states that may otherwise escape the primary LHC experiments undetected.  For example, the FASER experiment~\cite{FASER:2018bac} was initially proposed in a 2017 theory paper~\cite{Feng:2017uoz}.  FASER, which has already begun installation,  
will search for long-lived particles, such as dark photons, that are produced at the primary ATLAS interaction point with high rapidity and then decay in the FASER detector to {e.g.} lepton pairs.  FASER is representative of a broad class of new LHC experiments that have been proposed to search for new dark-sector or long-lived states far away from the primary interaction points, including MilliQan~\cite{Haas:2014dda}, MATHUSLA~\cite{Chou:2016lxi}, and CODEX-b~\cite{Gligorov:2017nwh}.

The LDMX experiment is another example of a novel detector concept that came out of theoretical work exploring possible ways of detecting feebly-coupled dark-sector particles.
In particular, theoretical work pointed out that dark sector models could be searched for by looking for their missing momentum signatures at fixed target experiments.  Smaller couplings with the SM could be probed since one does not have to both produce and detect the dark sector particles~\cite{Izaguirre:2014bca}.  The missing momentum approach to dark sector searches became the basis of the LDMX experimental proposal~\cite{LDMX:2018cma}.

Above, we enumerate a number of new accelerator-based detectors that arose from theoretical proposals for probing motivated dark sector parameter space, and that of long-lived particles more generally. However, in the past decade it has also been increasingly understood, thanks to theoretical efforts, that ATLAS, CMS, and LHCb may also be sensitive to dark sector paradigms using novel analyses~\cite{Essig:2022yzw}.  Searches of interest in dark sector models span a wide range, from light BSM states produced in meson decays to spectacular high-multiplicity events. 
While the connections between collider signatures and early universe dynamics are highly
model-dependent, high-multiplicity and/or displaced signatures constitute a powerful if technically challenging discovery opportunity at colliders and have become a growing focus of interest \cite{Albouy:2022cin}.  One major driver of theoretical work in this area is the need to assess current trigger coverage at the main LHC detectors and, critically, inform future trigger strategies, in order to maximize the chance that new physics produced at the LHC will indeed be detected, no matter how unconventional its signatures appear.

Broadly speaking, these examples illustrate how over the past decade theoretical constructions of dark sector DM models have caused theorists to propose novel laboratory-based experiments and analyses (in this case, accelerator experiments), which have become one of the central focuses of the high-energy experiment community today.

\section{Data-Driven Cosmology}

Cosmological observations offer the unique opportunity to reconstruct the history of the universe and the laws that shaped it.  We map the universe in frequency and angle (for both light and gravitational waves) from the local universe to the farthest cosmological distances.  It is through our theoretical understanding of the forces that shaped the cosmos~\cite{Amin:2022soj} that we can then reconstruct the expansion history, search for new particles and forces, identify new objects, and more.

\subsection{Cosmic Microwave Background}

The cosmic microwave background has been the main driving force in cosmology to date.  It provides the most precise measurement of cosmological parameters, both in $\Lambda$CDM and many extensions thereof.  Importantly, with polarization data from Planck~\cite{Planck:2018vyg}, these constraints are now robust to multi-parameter extensions of the cosmological Standard Model.  As CMB detector noise levels continue their Moore's law-type improvement~\cite{Chang:2022tzj,Abazajian:2022nyh,CMB-HD:2022bsz}, the impact of CMB data will depend increasingly on theoretical techniques to isolated different physical effects that alter these maps.  CMB photons are both gravitationally lensed~\cite{Lewis:2006fu} and scattered~\cite{Carlstrom:2002na} by the matter between us and the surface of last scattering.  Ongoing theoretical work has shown how these effects can be removed from the CMB maps using their statistical properties and frequency dependence.  In the process, we create new maps for the distribution of matter in the universe and the locations of high redshift galaxy clusters.  These maps can themselves contain valuable information about the history of the universe and fundamental physics.  For example, the lensing of the CMB is expected to play an essential role in detection of a non-zero sum of the neutrino masses~\cite{Dvorkin:2019jgs} and other light (but massive) particles~\cite{Dvorkin:2022jyg} with the next generation of cosmic surveys.  CMB bounds on light relics already provide the leading constraint on many reference models of feebly-interacting particles in substantial portions of parameter space (see, e.g.~\cite{Agrawal:2021dbo}).

Galactic foregrounds, particularly from dust, present an additional challenge to CMB measurements, particularly for the measurement of primordial gravitational waves via the polarization B-modes~\cite{BICEP2:2015nss,BICEP:2021xfz}.  Polarized dust emission in our galaxy produces both E-modes and B-modes~\cite{Planck:2018gnk}.  The amplitude of the B-modes from dust, at the peak CMB frequencies, is larger than the expected primordial signal and must be removed from the maps in order to constrain or detect gravitational waves from inflation~\cite{CMB-S4:2020lpa}.  Our understanding of these dust signals from first principles is limited but has been bolstered by simulations~\cite{Kim:2019xov,Kritsuk:2017aab} and data-driven techniques~\cite{Thorne:2021nux,Krachmalnicoff:2020rln}.  Continued development of these techniques will be essential in order to reach the sensitivity goals of the next generation surveys like the Simons Observatory~\cite{SimonsObservatory:2018koc}, CMB-S4~\cite{Abazajian:2022nyh}, and CMB-HD~\cite{CMB-HD:2022bsz}.

\subsection{Gravitational Wave Observatories}

Gravitational waves produced after inflation are an increasingly compelling window into the history of the universe. This includes both gravitational waves produced during reheating and/or phase transitions in the early universe~\cite{Caldwell:2022qsj} and from mergers of black holes, neutron stars, and/or more exotic objects~\cite{Berti:2022wzk}. Given that the universe is transparent to gravitational waves, this presents a unique opportunity to probe important events throughout cosmic history and not just after recombination.

One of the guaranteed signals in the coming generation of observations is the measurement of the expansion history via the gravitational wave signal from binary inspirals~\cite{Berti:2022wzk}. The increasing sensitivity of surveys will allow for the discovery of many more binaries at much larger distances.  However, with this increased sensitivity comes much higher demands of accuracy for the inspiral templates.  Generating templates at the accuracy needed for future observatories will require a variety of theoretical tools, in addition to simulations~\cite{Lehner:2014asa}, that connect observational needs to the most fundamental questions about the nature of gravity.  The current state of the art for analytic calculations has been pushed forward by a combination of EFT~\cite{Porto:2016pyg} and amplitudes techniques~\cite{Buonanno:2022pgc}. In addition, these calculations expose unusual properties of black hole Love numbers~\cite{Porto:2016zng}, which has driven purely theoretical investigations into the symmetries of horizons.  The calculation of the waveform from amplitudes techniques also benefits and exposes properties of the double copy~\cite{Adamo:2022dcm}.

\subsection{Large Scale Structure}

The distribution of galaxies at lower redshifts  is an increasingly important window into the universe~\cite{Ferraro:2022cmj,Dawson:2022oig,Karkare:2022bai,Liu:2022iyy}.  In many ways, the raw statistical power of large scale structure surveys rivals the CMB today and will rapidly exceed it with surveys like DESI~\cite{DESI:2016fyo} and Euclid~\cite{Amendola:2016saw}.  Unfortunately, our ability to use these maps to understand the universe is limited by our understanding of nonlinear structure formation and astrophysical uncertainties, not simply statistics. Theoretical insights paired with increasingly sophisticated simulations have made many new and more powerful analyses possible~\cite{Amin:2022soj}, but much work remains if we are to harness the full power of these surveys for fundamental physics.

On the largest scales, EFT~\cite{McDonald:2009dh,Baumann:2010tm,Carrasco:2012cv} and perturbation theory~\cite{Bernardeau:2001qr} techniques are sufficient to model both the DM, baryons and tracers thereof.  Work on the effective theory of large scale structure has made the regime of validity transparent, along with the systematic expansion in the density, and has opened the door to analyses of the full-shape matter power spectrum~
\cite{Ivanov:2019pdj,DAmico:2019fhj} and bispectrum~\cite{Cabass:2022wjy,DAmico:2022gki} that have never previously been done (and were first performed by theorists).  Unlike in the CMB, EFT techniques~\cite{Cabass:2022avo} (including the bias expansion for galaxies~\cite{McDonald:2009dh}) are essential as the microphysics is not known in detail.  One of the fundamental principles of EFT is that unknown short distance physics can be absorbed into the parameters of the EFT, allowing us to fit the EFT parameters to data without the need to simulate this short distance physics perfectly.  Many signals of interest can also be cast into an EFT framework~\cite{Cabass:2022avo, Boddy:2022knd} so that both the fundamental and astrophysical components of the data can be measured simultaneously. 

In addition to modeling the LSS perturbatively, theoretical understanding of both the signal and the nonlinear evolution have yielded a number of protected observables, quantities that we can measure that are largely not impacted by nonlinear evolution.  The BAO are one of the most significant examples: although the oscillatory feature appears at the nonlinear scale in the matter power spectrum, it corresponds to physics at the distance of the sound horizon at recombination~\cite{Eisenstein:2006nj}.  Since the DM moves slowly, its nonlinear evolution cannot mimic this large scale effect. For this reason, measurements of the BAO have been demonstrated to be highly robust to nonlinearity, yet can include modes that would typically be omitted from a power spectrum analysis~\cite{BOSS:2016hvq}.  Further work has revealed that the phase of the BAO oscillation is also protected from nonlinear corrections~\cite{Baumann:2017lmt}, enabling a new constraint on $N_{\rm eff}$ from LSS alone~\cite{Baumann:2019keh}.  Furthermore, additional oscillatory signals from inflation can also be searched for in the nonlinear regime for the same reason and give rise to the best constraint on primordial features to date~\cite{Beutler:2019ojk}.  Naturally, the special features of the BAO are not limited to the power spectrum, and oscillatory higher point functions related to the BAO~\cite{Slepian:2014dda} or inflation~\cite{Flauger:2010ja} can also be analysed in the same way~\cite{Slepian:2016kfz}.

Searches for primordial non-Gaussianity~\cite{Achucarro:2022qrl} in large scale structure are protected from nonlinear evolution for their own distinct reasons~\cite{Alvarez:2014vva}.  Local non-Gaussianity, parameterized by $\fnlloc$, arises in multi-field inflation (more than one nearly light field), and produces a scalar metric fluctuation 
\beq
\Phi(\x) = \varphi_g(\x) - \fnlloc \varphi_g^2(\x) + \ldots \ ,
\eeq
where $\Phi(\x)$ is the Newtonian potential and $\varphi_g(\x)$ is a Gaussian random field.  The resulting statistics violate the signal-field consistency conditions~\cite{Maldacena:2002vr,Creminelli:2004yq}.  More dramatically, it was shown by theoretical arguments and validated with simulations that these non-Gaussianity statistics give rise to scale dependent bias~\cite{Dalal:2007cu}, namely the density contrast of galaxies $\delta_g(\x) \propto \fnlloc \Phi(\x)$. As $\Phi$ is unphysical in Newtonian gravity (by the equivalence principle), this signal cannot arise from short distance nonlinearities.  These implications can be generalized as a set of large scale structure consistency conditions~\cite{Creminelli:2013mca,Creminelli:2013poa}.  Even additional heavy fields that are produced during inflation will violate these conditions, providing an avenue to measure the spectrum of particles present during inflation with large scale structure surveys~\cite{Chen:2009zp}.  New types of protection relevant to other forms of non-Gaussianity may emerge at the level of the maps of LSS to enforce local and causal evolution.

These insights into the nature of the cosmological signals have driven major progress in analyzing data from surveys. For illustration, the BAO feature is famously sensitive to large-scale velocity effects.  In current analyses, these bulk effects are measured and removed using BAO reconstruction~\cite{Eisenstein:2006nk}. Similarly, for primordial non-Gaussianity, our understanding of scale-dependent bias alone is expected to provide the dominant constraint on local non-Gaussianity~\cite{Dore:2014cca}.  Our understanding of this signal also led to the development of cosmic variance cancellation~\cite{Seljak:2008xr} which is likely to be an essential tool in upcoming surveys~\cite{Schmittfull:2017ffw}.

The continued understanding of cosmological data in the context of BSM physics is important for developing new strategies for analyzing cosmological data. The impact of DM around the time of recombination continues to inform new analyses of cosmological data including the effects of annihilation, decay and interactions with baryons.  In addition, light axion-like particles produce a wide range of effects on cosmological scales, including modulating the time series data from CMB polarization~\cite{Fedderke:2019ajk}.  These theoretical insights expand the scientific impact of existing surveys and shape the goals and design of future projects.

\subsection{Small Scale Structure}

Over the next decade astrophysical surveys will provide an unprecedented amount of information about the DM distribution on galactic and sub-galactic scales \cite{Boddy:2022knd, Banerjee:2022qcb}.  These observations will substantially expand the census of dwarf galaxies and have the potential to detect DM-only halos through their gravitational interactions, {\it e.g.}~through strong gravitational lensing or the disruption of stellar streams.  
The small-scale cutoff in the matter power spectrum depends on the particle properties and interactions of DM, and new observations sensitive to the small-scale matter power spectrum have the potential to provide powerful and possibly conclusive tests of the cold DM hypothesis.  
Large- and small-scale simulations of gravitational dynamics, essential to understanding cosmological data in many scenarios, will be particularly critical to translate the forthcoming wealth of galactic-scale observations into reliable inferences about DM properties \cite{Alvarez:2022rbk,Boddy:2022knd}.  

On small scales, the density contrast becomes order one and perturbative methods break down.  In addition, on these nonlinear scales baryonic feedback significantly alters both the distribution of DM and the formation of galaxies.  DM-only simulations have been important in understanding this regime, but have left a number of outstanding problems concerning the distribution of matter on small scales, including the DM at cores of galaxies and the number of satellites.  Simulations of DM and baryons, including feedback from active galactic nuclei and supernovae, have suggested that baryons could play a significant role in addressing these problems~\cite{Vogelsberger:2019ynw}.  In contrast, a number of models of DM~\cite{Asadi:2022njl} have been proposed that would alleviate these tensions between simulations and observations, by introducing small-scale cutoffs in the matter power spectrum, giving DM self-interactions, or both~\cite{Bechtol:2022koa}.
Meanwhile, for many of the novel observational avenues proposed to probe small-scale DM structure using the Vera C. Rubin Observatory, JWST, and the Nancy Grace Roman Space Telescope, novel hydronamical simulations are necessary in order to establish clear predictions; for many of these types of probes such simulations are still at very early stages of development \cite{Banerjee:2022qcb}.  

Continued development of simulations on both cosmological and galactic scales are therefore essential to our understanding of the fundamental physics of DM~\cite{Banerjee:2022qcb, Boddy:2022knd} and the astrophysics of galaxies and galaxy clusters.  This will require not only advances in modeling baryons within cosmological simulations, but also in simulating halo formation in models of DM with astrophysically-important particle properties.

DM with sufficiently complex particle interactions may have even more dramatic effects on structure within galaxies, particularly if this DM makes up only a fraction of the DM of our universe.  In parts of model space this DM may form collapsed dark structures, whether through dissipative dynamics, condensation, or as primordial BHs \cite{Asadi:2022njl, Bechtol:2022koa,Brito:2022lmd, Bird:2022wvk}. The typical scales and distributions of the resulting objects are highly model-dependent, but offer new and distinct avenues for gravitational probes, whether through gravitational waves, extended dark structures such as miniclusters, or dark compact objects. 

At very small scales (Earth mass and below), corresponding to scales that entered the horizon prior to BBN, a variety of mechanisms can lead to a dramatically increased abundance of DM microhalos \cite{Bechtol:2022koa}.  Departures from radiation domination prior to BBN, e.g.~an early matter domination era (EMDE), can give rise to substantial enhancements in primordial perturbation growth on scales that experience this modified expansion, up to a small-scale cut off that depends on the microphysics governing both DM and the particles responsible for generating the modified expansion \cite{Erickcek:2011us,Redmond:2018xty}.  
Provided that density perturbations remain linear during the EMDE itself, and that DM remains sufficiently cold following the EMDE, this period of rapid perturbation growth can give rise to enhanced small-scale structure in the late universe \cite{Bechtol:2022koa}.  The microhalos that form out of the resulting density field collapse much earlier than they would in a standard $\Lambda$CDM cosmology, resulting in substantially denser structures that can persist within galactic environments today. 

Nonthermal DM dynamics in the early universe can also lead to similar kinds of enhanced DM microstructures.  Axion minihalos are perhaps the primary example.  Enhanced small-scale structure can also arise in theories where DM is produced purely gravitationally by stochastic fluctuations 
during inflation.  This inflationary production of DM results in DM isocurvature perturbations, which are notably constrained on the scales probed by the CMB\null.  These constraints are challenging for the gravitational production of scalar DM, but unlike for scalars the inflationary fluctuations in (longitudinal) vector fields are not scale-invariant: the dimensionless power spectrum instead goes like $\Delta^2\sim (k H_I/m)^2$, which is peaked at the scale $k\sim a_m m$, where $a_m$ is the scale factor for which $H(a_m) = m$ \cite{Bechtol:2022koa}.  This matter power spectrum is also enhanced on small scales compared to $\Lambda$CDM and similarly gives rise to an enhanced population of early-forming minihalos.

If DM has an annihilation cross-section into visible particles, then having an  $\mathcal{O}(1)$ fraction of DM bound into these dense microhalos provides an enormous boost to the annihilation rate, allowing stringent constraints to  be set on these scenarios already from Fermi observations \cite{Bechtol:2022koa}.  Discovering and characterizing an enhanced microhalo population using purely gravitational interactions is more futuristic. Some promising observational avenues are pulsar timing arrays and cluster caustic microlensing \cite{Bechtol:2022koa, Blinov:2022tfy}.   
Although observational capabilities for both of these signatures are set to dramatically improve over the next decade, substantial theoretical work remains to be done to solidify the observational prospects for these scenarios.

\subsection{The Next Decade}

We have only one universe to observe and making the most of these observations requires the continued development of theoretical techniques.  As current observations are increasingly limited by our theoretical models and data analysis techniques, rather than the raw noise levels of maps themselves, there is a growing opportunity for theorists to make an enormous impact on our understanding of the universe from these data. Amusingly,~\cite{McDonald:2009dh} attempted to quantify this impact in terms of dollars.  For most cosmic parameters, the constraining power of a survey is determined by the number of linear Fourier modes we can reconstruct.  For example, in a three-dimensional survey the number of modes is roughly $N_{\rm modes} \approx V k^3_{\rm max}$ where $V$ is the survey volume and $k_{\rm max}$ is the maximum wavenumber one can realistically model.  The current generation of CMB and LSS surveys measure approximately $10^6$ modes at roughly 1000 USD per mode.  Yet, if improvements in theoretical modeling of the data can increase $k_{\max}$ by even 30\%, it effectively doubles the sensitivity of the survey.  In this precise sense, a small investment in theoretical cosmology can yield billions of dollars of scientific value.  

The past few years have seen significant advances in modeling and analysis techniques across the spectrum of cosmological observables.  These insights have already produced meaningful improvements over conventional techniques.  In the next decade these developments are expected to be transformative, enabling orders of magnitude improvements in some cases.  Theoretical techniques are increasingly an equal component of progress in observational cosmology and demand future investments that reflects this impact.

Meanwhile on sub-galactic scales our understanding of the distribution of matter has rapidly developed in recent years and is poised to expand yet further with the advent of new facilities \cite{Brito:2022lmd,Bechtol:2022koa, Chakrabarti:2022cbu,Ballmer:2022uxx}.
While there are many exciting opportunities here to detect particle properties of DM that would otherwise be out of reach, reaching the full scientific potential of this widened view of the cosmos will require a substantial investment in understanding the evolution of both known and as-yet-theoretical components of our universe deep in the nonlinear regime.

\section{Cosmology Meets Fundamental Theory}

The expansion of the universe is accelerating today and was very likely accelerating in the past, due to dark energy and inflation, respectively.  These epochs present unique challenges for fundamental physics, both qualitatively and quantitatively.  Recent progress has been driven by a variety of advances connecting cosmology to the many corners of the theory frontier~\cite{Flauger:2022hie}.  Yet, understanding quantum gravity in cosmological spacetimes remains one of the largest and most important unsolved problems in high energy physics, as it unites both basic theoretical questions, cosmological observations, and even the origin and fate of the universe.

\subsection{Inflation}

Inflationary cosmology provides a controlled context in which to build our theoretical understanding of accelerating spacetimes.  This is due to the fact that inflation ends, leading to a natural physical regulator of the long distance behavior.  Yet, more importantly, the end of inflation with reheating and a hot big bang implies that quantum fluctuations produced during inflation are ultimately observable by the inhabitants of our universe.  Understanding how to calculate these signals provides a test of our theoretical understanding of the inflationary epoch~\cite{Flauger:2022hie,Baumann:2022jpr} and an opportunity to test inflation with current and future cosmological surveys~\cite{Achucarro:2022qrl}.

In recent years, our understanding of inflation has been significantly enhanced by the insights of EFT~\cite{Cabass:2022avo}.  An EFT for the fluctuations, the EFT of Inflation~\cite{Cheung:2007st}, gives a model independent definition of inflation.  Inflation is characterized by two conditions (a) a period of near exponential expansion such that $|\dot H(t)| \ll H^2(t)$ and (b) a physical clock that breaks the microscopic time-translation symmetry, $\langle {\cal O}\rangle \propto t$ (which is ultimately gauged by gravity).  The first condition ensures that long wavelength fluctuations are produced from cosmological particle production while the second is needed to end the inflationary epoch.  This pattern of symmetry breaking gives rise to a goldstone boson $\pi$ that is eaten by the metric to yield the scalar metric fluctuation $\zeta = - \pi H+{\cal O}(\pi^2)$.  The predictions for inflation are encoded in the equal-time in-in~\cite{Maldacena:2002vr,Weinberg:2005vy} correlators of the scalar ($\zeta$) and tensor metric fluctuations ($\gamma_{ij}$), e.g.
\beq
\langle \zeta(\k_1,t) .. \zeta(\k_n,t) \rangle = F_n(\k_1,..,\k_n) (2\pi)^3 \delta(\sum_i \k_i ) \ ,
\eeq
where $\k_i$ are the comoving momenta.  The shape~\cite{Babich:2004gb} and amplitude of these correlation functions is determined by the particle content and interactions in this EFT\null. Because of the intrinsically nonlinear nature of gravity, there is a lower bound of the size of these correlators \cite{Cabass:2016cgp}, which lies two orders of magnitude below the current sensitivity.

In many models of inflation, the role of the clock is played by a scalar field $\phi$ rolling on a potential $V(\phi)$ such that $\phi \propto t$.  However, observations demand that this potential is extremely flat; concretely, we require that $\eta 
\equiv M_{\rm pl}^2 V''/V \ll 1$.  We see that Planck suppressed dimension six operators of the form $V(\phi) \to V(\phi)(1+ \phi^2/(2\Mpl^2)) $ shift $\eta \to \eta+1$, making the problem of building a model of inflation potentially sensitive to physics at the Planck scale, regardless of the scale of the inflationary potential.  Known as the $\eta$-problem, this observation has motivated the need to understand models in inflation in the context of string theory~\cite{Baumann:2014nda}, or to look for field theory mechanisms to eliminate this sensitivity. 

In addition to scalar fluctuations, primordial gravitational waves are also produced during inflation with an amplitude that is typically tied to the Hubble scale during inflation, $\Delta_T = 4 H^2/ \Mpl^2$.  Furthermore, the amplitude of the scalar power spectrum determines the speed of the rolling scalar field, $|\dot \phi| = (59 H)^2$, and hence the field range 
\beq
\frac{\Delta \phi}{M_{\mathrm{pl}}} \gtrsim\left(\frac{r}{8}\right)^{1 / 2} N_{\star} \gtrsim\left(\frac{r}{0.01}\right)^{1 / 2} \ ,
\eeq
where $r = \Delta_T / \Delta_\zeta$ is the tensor to scalar ratio and $N_\star \geq 30$ is the number of e-folds of inflation.  For $r$ in range of future measurements, $r > 10^{-3}$, the field range is also super-Planckian~\cite{Lyth:1996im}.  Control of the potential over such large distances requires more than forbidding a few irrelevant operators, but requires some approximate symmetry of the UV completion at the Planck scale.

Inflationary model building makes extreme demands of the EFTs that describe the inflationary mechanism.  This has inspired a wide range of work investigating the space of consistent EFTs~\cite{deRham:2022hpx} and their origins from a UV complete theory like string theory~\cite{Draper:2022pvk}.  These investigations have also enriched the landscape of inflationary phenomenology, as the search for explicit models has yielded a variety of novel ideas~\cite{Alishahiha:2004eh,Silverstein:2008sg,McAllister:2008hb,Agrawal:2022rqd} that can be studied and generalized in the context of inflationary EFTs~\cite{Cabass:2022avo}, and yield new signals and analyses of cosmological data~\cite{Achucarro:2022qrl}.

In recent years, the constraints of self-consistency and the existence of a UV completion have been applied directly to the EFT of inflation and the cosmological correlators.  Often characterized as the cosmological bootstrap~\cite{Baumann:2022jpr}, this approach shares common elements with the amplitudes~\cite{Kruczenski:2022lot} and conformal bootstrap programs~\cite{Hartman:2022zik}.  These connections are most transparent in nearly de Sitter backgrounds, where significant progress has been made understanding the space of cosmological correlators from first principles.  Yet, the ultimate ambition of this program is to understand the full space of self-consistent inflationary models and their predictions.  In some cases, the challenge is translating existing bounds from flat space and anti-de Sitter (AdS) to de Sitter (dS) space.  However, as the inflationary background breaks Lorentz invariance, the natural questions about inflationary EFTs can inspire new questions in AdS and flat space as well~\cite{Baumann:2019ghk}.

One promising opportunity at the intersection of fundamental physics and cosmological observations is the impact of the particle spectrum during inflation, including light and heavy fields, on non-Gaussian cosmological correlators~\cite{Chen:2009zp}.  Every particle is produced from the vacuum during inflation, albeit with an exponentially suppressed amplitude for $m\gg H$ (e.g.~$m > 3 H/2$ for scalars).  Once produced, these particles can alter the fluctuations of the inflaton itself and the resulting correlators.  The spectra and interactions of such particles is a window into the underlying theory at energies that are potentially near the GUT scale.  In addition, these observational signatures are fundamentally distinct from single field inflation~\cite{Creminelli:2004yq} and robust to other details of the inflationary model.

Interestingly, the tail of the distribution of the scalar fluctuations is particularly sensitive to the specific UV model (see e.g.~\cite{Panagopoulos:2019ail,Panagopoulos:2020sxp,Celoria:2021vjw}). Understanding the non-perturbative nature of these large fluctuations is a theoretical problem that connects the more conceptual challenges of inflation, such as eternal inflation, with observational questions like the formation of primordial black holes~\cite{Carr:2016drx,Vennin:2020kng,Bird:2022wvk}. The novel techniques that will be needed to better understand rare fluctuations will likely have impact far beyond the cosmological domain.  

\subsection{Dark Energy and de Sitter Space}

The discovery of the accelerated expansion of the universe at late times further challenges our understanding of fundamental physics.  The small size of the cosmological constant has resisted a natural explanation~\cite{Weinberg:1988cp}.  In addition, evidence from string constructions and other theoretical challenges have suggested that perhaps de Sitter space is fundamentally unstable and/or is one of many accelerating regions~\cite{Polchinski:2006gy}.  It remains entirely possible that our current acceleration is more analogous to a second inflationary period, driven by a more dynamical form of dark energy, and that de Sitter-like regions do not occur at all.  Observational constraints on this possibility are encoded in the EFT of dark energy that is constrained by a variety of observational and gravitational probes~\cite{Cabass:2022avo}.

The challenges associated with late-time acceleration range from perturbation theory on a fixed background all the way to non-perturbative effects in a quantum theory of gravity.  Arguably, de Sitter space is the best testing ground for these questions about dark energy, as it is both the most well-studied accelerating cosmology and fits current observations. In perturbation theory, fluctuations in dS contain IR divergences and secular terms whose interpretations and resolutions have been debated over the years.  Non-perturbatively, it is not even clear what a quantum theory of de Sitter space should even compute.  These issues are connected, as they originate from the fact that all fields, including the metric, fluctuate.  Given an infinite amount of time, the amplitude of the scalar fluctuations will diverge somewhere.  

Recently, significant progress has been made understanding the structure of perturbation theory in a fixed de Sitter background~\cite{Gorbenko:2019rza,Mirbabayi:2019qtx,Baumgart:2019clc,Cohen:2020php}.  Using EFT or diagrammatic techniques, one can see both the IR and secular terms that arise in the cosmological slicing of dS can be resummed using the framework of stochastic inflation.  Some higher loop divergences introduce corrections to these equations that can be solved to give non-perturbative results for the probability distributions of the fields and their correlators.

Progress understanding perturbative quantum gravity in de Sitter has helped sharpen the non-perturbative challenge of de Sitter quantum gravity.  One of the central questions about the nature of de Sitter is the interpretation of the Gibbons-Hawking entropy associated with the cosmological horizon.  Recent progress in black hole physics~\cite{Bousso:2022ntt} has provided increasingly compelling evidence that the black hole behaves as a quantum system with $e^S$ states, where $S$ is the Bekenstein-Hawking entropy.  While it is natural to associate a finite number of degrees of freedom to the cosmological horizon, it is less clear how to interpret such a statement.  Yet, it was argued in~\cite{Arkani-Hamed:2007ryv} that such an interpretation would preclude the possibility of seeing more than $e^S$ fourier modes after inflation ends.  It has been verified in examples, away from the eternally inflating regime~\cite{Creminelli:2008es}, but is sensitive to the statistical properties of scalar fluctuations in de Sitter~\cite{Cohen:2021jbo}.  This provides a sharp connection between inflationary model building, perturbative calculations and the physics of de Sitter space.  In addition, non-perturbative instabilities will also lead to eternal inflation and thus add further complexity to this picture~\cite{Freivogel:2005vv}.

It has long been hoped that de Sitter holography could clarify some of the questions, as it did in AdS\null.  However, there is no fixed boundary on which to anchor observables in dS and defining de Sitter observables remains an unsolved problem. A variety of approaches attempt to circumvent these issues in different ways. For example, dS/CFT~\cite{Strominger:2001pn,Maldacena:2002vr} is a holographic theory for the wavefunction for a given fixed boundary metric.  Cosmological correlators are then computed by integrating over the metric weighted by the probability distribution~\cite{Maldacena:2002vr}.  This approach is directly connected to traditional cosmological correlators and provides a useful perspective on the symmetries of the inflationary fluctuations~\cite{Baumann:2022jpr}.  In contrast, dS/dS~\cite{Alishahiha:2004md} is a holographic approach where the boundary theory also includes dynamical gravity and thus does not reduce to a field theory in the sense of AdS/CFT\null.  Nevertheless, progress in dS/dS has connected to aspects of string model building and the entropy of de Sitter. More generally, a number of perspectives on de Sitter holography  have emerged in recent years.  Ultimately, one might hope that one, several, or all of these approaches will shed light on the fate of the universe, the space of inflationary models, and/or the (non-perturbative) meaning of the wavefunction of the universe.

\subsection{The Next Decade}

The understanding of cosmological backgrounds in perturbation theory has seen major progress over the past decade.  Bootstrap methods have dramatically expanded our understanding of cosmological correlators at tree level.  EFT techniques have helped clarify and organize loop calculations to the point where the long distance and late time behavior can be controlled in most examples.  This progress is being spurred by rapid developments across the theory frontier, including the conformal bootstrap and the amplitudes program in particular.

Despite this progress, the most tantalizing questions in cosmology require non-perturbative insights that have remained elusive.  We increasingly understand how to calculate correlators in a given inflationary model; yet, characterizing the space of consistent inflationary models and their observational signatures remains a vastly more difficult problem to solve.  We can similarly calculate the wavefunction of the universe for small fluctuations around a fixed background, but struggle to understand the wavefunction for comparing different backgrounds or large fluctuations.  

Recent progress in understanding the quantum nature of black holes~\cite{Bousso:2022ntt} offers hope that answering these longstanding cosmological questions may soon be within reach.  Furthermore, progress driven by the cosmological bootstrap program and cosmological collider physics has begun to pinpoint the uniquely quantum aspects of inflationary predictions~\cite{Green:2020whw} and might draw further connections to the tools of quantum information theory that have proved powerful for QFT~\cite{Faulkner:2022mlp} and black holes. Every scale in physics intersects our cosmological history and therefore exposing the secrets of the universe may naturally require insights from across the theoretical landscape.

\section{Outlook and Conclusions}\label{sec:conclusions}

Theoretical particle astrophysics and cosmology are increasingly central parts of the quest to measure and understand the universe, as the line between theory and experiment is blurred in the context of cosmological and astrophysical tests of high energy physics. Theorists have been essential in the conceptual design of a number of ambitious experiments, the definition of the critical scientific targets, the development of novel analyses of the data and the connection to the fundamental laws of nature.  It is essential that investments in the coming decade embrace the expanding role of theory to broaden the scientific goals and impact of the cosmic frontier within high energy physics and beyond.

Simultaneously, progress answering many profound questions that we have about our universe is inseparable from progress in the rest of the theory frontier. Theoretical tools developed for a variety of reasons come to bear on questions about the nature of DM, dark energy and inflation.  Furthermore, a deeper understanding of a variety of quantum systems expands our ability to use new devices to search for signals of cosmic relics. Significantly, our ability to analyze and interpret cosmic data is often limited by our ability to calculate and understand astrophysical signals, which have been improved by both analytic and computational techniques.  

This Report outlined a number of critical areas and goals for the coming decade.  For the physics of DM, the coming decade is likely to bring new methods for the direct detection of DM, including axion DM, and especially utilizing emerging quantum technologies.  We anticipate close collaboration between theorists and experimentalists bringing these ideas to reality.  The next decade is also likely to bring new advances in our understanding of DM's possible non-gravitational interactions through cosmological and astrophysical observations in systems ranging from our own galaxy to the CMB\null.  Here too, close collaboration between theorists and observers will be important for maximizing the scientific return from current and planned observations.
In parallel to observational advances, we can also anticipate theoretical advances developing new models for DM that will be tested by these observations, including dark sectors and theories that connect to broader frameworks like the hierarchy problem.

The coming decade in theoretical cosmology is likely to include significant challenges in modeling and isolating astrophysical foregrounds from key theoretical targets. This is particularly well known in the cases of dust for primordial gravitational waves (CMB B-modes), and baryonic physics for measuring the initial density fluctuations with galaxy surveys (primordial non-Gaussianity).  Yet, deepening our understanding of the history of the universe is a much broader goal covering a wide range of astrophysical and cosmological epochs. Insights into the dynamics of the universe on cosmological scales has consistently revealed new opportunities to test fundamental physics in the sky. Novel contributions from theory, simulation, and data analysis will be essential for unlocking the full potential of the next generation of cosmic surveys.

Questions about the nature of the universe as a whole remain central to the goals of the theoretical physics community.  Astrophysical and cosmological observables offer a unique glimpse into high energy processes far beyond those accessible on Earth, and offer a window into the most basic structures that define the fundamental laws.  In addition, the inflationary epoch and currently accelerating universe raise profound questions about the quantum nature of the universe, its origin and future.  The full breadth and depth of theoretical physics is intertwined with our understanding of the universe and displays the richness of the theory frontier as a whole.

\paragraph{Acknowledgements}

\hskip 10pt DG was supported by the US~Department of Energy under Grant \mbox{DE-SC0009919}.  JTR was supported by NSF grant PHY-19154099 and NSF CAREER grant PHY-1554858.  BRS was  supported   in   part   by   the   US~Department of Energy   Early   Career   Grant   \mbox{DE-SC0019225}. The work of JFS was supported in part by the US~Department of Energy Early Career Grant \mbox{DE-SC0017840}.

\clearpage
\phantomsection
\addcontentsline{toc}{section}{References}
\bibliographystyle{utphys}
\bibliography{TF09refs}

\end{document}